\documentclass[amsfonts,letterpaper,aps,prb,preprint,groupedaddress,citeautoscript]{revtex4}

\usepackage{graphicx} %for figures
\usepackage{amsmath}
\usepackage{bm} % for some bold math symbols
\begin{document}

\title{Excitonic effects on the two-color coherent control of
interband transitions in bulk semiconductors}

\author{R. D. R. Bhat}
\author{J. E. Sipe}
\affiliation{Department of Physics and Institute for Optical
Sciences, University of Toronto, 60 St. George Street, Toronto,
Ontario M5S 1A7, Canada}

\date{\today}

\begin{abstract}
Quantum interference between one- and two-photon absorption pathways
allows coherent control of interband transitions in unbiased bulk
semiconductors; carrier population, carrier spin polarization,
photocurrent injection, and spin current injection may all be
controlled. We extend the theory of these processes to include the
electron-hole interaction. Our focus is on photon energies that
excite carriers above the band edge, but close enough to it so that
transition amplitudes based on low order expansions in $\mathbf{k}$
are applicable; both allowed-allowed and allowed-forbidden
two-photon transition amplitudes are included. Analytic solutions
are obtained using the effective mass theory of Wannier excitons;
degenerate bands are accounted for, but envelope-hole coupling is
neglected. We find a Coulomb enhancement of two-color coherent
control process, and relate it to the Coulomb enhancements of one-
and two-photon absorption. In addition, we find a frequency
dependent phase shift in the dependence of photocurrent and spin
current on the optical phases. The phase shift decreases
monotonically from $\pi /2$ at the band edge to $0$ over an energy
range governed by the exciton binding energy. It is the difference
between the partial wave phase shifts of the electron-hole envelope
function reached by one- and two-photon pathways.
\end{abstract}


\maketitle
\section{Introduction\label{sec:introduction}}

The phenomenon of quantum interference can be used to control
physical and chemical processes. One approach, the `$n+m$' scheme,
uses a two-color light field to interfere $n$- and $m$-photon
transitions \cite{Manykin67,ShapiroBrumer97,Gordon99}. Interference
between one- and two-photon transitions, for example, allows
controllable polar asymmetry of photoelectrons in atomic ionization
\cite{Yin92,Baranova92}, controllable dissociation of HD$^{+}$
\cite{Sheehy95}, and controllable photocurrent injection in unbiased
solids due to free carrier absorption \cite{Baskin88}, impurity-band
transitions \cite{Entin89}, and quantum well bound-continuum
intersubband transitions \cite{Dupont95}. In biased asymmetric
semiconductor double wells, `$1+2$' interference allows control of
carrier population and THz emission \cite{Potz98}. Interband `$1+2$'
interference in unbiased semiconductors, which is our interest here,
allows independent control of electrical current injection
\cite{Atanasov96,Hache97} and spin current injection
\cite{BhatSipe00,StevensJAP,StevensPRL03,HubnerPRL03,Najmaie03,Marti04}.
Furthermore, in noncentrosymmetric semiconductors, it allows
independent control of carrier populations (i.e., absorption)
\cite{Fraser99PRL,Fraser03} and carrier spin polarization
\cite{Stevens_pssb, StevensReview04}. In each scenario, the
experimenter can control the interference by adjusting the phases of
the two colors.

The controllable optical phases are not the only source of phase
between the transition amplitudes. In general, there is also a
material-dependent intrinsic phase \cite{Shapiro88}.
Phenomenologically, the intrinsic phase appears as a phase shift in
the dependence of the process on a relative phase parameter of the
optical fields. Additionally, selectivity between two processes is
possible when their intrinsic phases differ
\cite{ShapiroBrumerBook}; for example, `1+3' experiments on diatomic
molecules have controlled the branching ratio of ionization and
dissociation channels \cite{Zhu95}. The intrinsic phase can be
strongly frequency dependent near resonances \cite{Zhu95}, and the
hope that it might serve as a new spectroscopic observable
\cite{Seideman98,Seideman99} has led to efforts to understand its
physical origin.

Whereas a resonance is necessary for a phase shift to a `1+3'
process \cite{Seideman98}, it is not necessary for a phase shift to
a `1+2' process. For example, an intrinsic phase in the `1+2'
photoionization of atoms is predicted from the simple model of a
delta function potential \cite{Baranova90, Pazdzersky97}.

Nevertheless, microscopic theories for the interband `$1+2$'
processes in bulk semiconductors have until now predicted trivial
intrinsic phases
\cite{Atanasov96,Fraser99PRL,BhatSipe00,Stevens_pssb}. The
photocurrent, for example, was predicted to be proportional to the
sine of the relative phase parameter for all final energies
\cite{Atanasov96}. Each of these theories use the independent
particle approximation, in which the Coulomb attraction between the
injected electron and hole is neglected. That approximation is
expected to be good for final energies well above the band gap,
since in this case the electron and hole travel quickly away from
each other. However, close to the band gap, one generally expects to
see deviations from the independent particle approximation. In the
one-photon absorption spectrum, for instance, it is well known that
the electron-hole attraction is responsible for exciton lines below
the band gap, and for an enhancement of the absorption above the gap
known as \textit{Sommerfeld} or \textit{Coulomb enhancement}
\cite{HaugKoch}.

The effect of the electron-hole attraction on one-photon absorption
has been studied with various degrees of sophistication. On the one
hand, modern \textit{ab initio} calculations that include Coulomb
effects have recently given very good quantitative agreement with
experimental spectra \cite{Albrecht98, BenedictPRL98, BenedictPRB98,
Rohlfing98, Rohlfing00}, although at the cost of significant
computational overhead. On the other hand, simple models of Wannier
excitons can describe Coulomb effects near the band edge of many
direct gap semiconductors. These excitonic effects have long been
understood qualitatively on the basis of a simple two-band model in
the effective mass approximation \cite{Elliott57}, which is even
quantitatively accurate for typical semiconductors
\cite{Rohlfing98}.

Excitonic effects on nonlinear optical properties of bulk
semiconductors have also been studied in the effective mass,
Wannier exciton approximation
\cite{Loudon62,Mahan68,Rustagi73,LeeFan74,Doni74,
Sondergeld77II,Girlanda81,Blossey70, Ganguly67, Martin71,
GarciaCristobal94, GarciaCristobal98, Kolber78, SheikBahae94} and
only recently with \textit{ab initio} methods \cite{Chang02}. The
two-photon absorption spectrum shows a different set of exciton
lines and a Coulomb enhancement that is weaker than its one-photon
counterpart.

One- and two-photon absorption spectra have been measured
sufficiently often that excitonic effects on them are well
established. In contrast, semiconductor `1+2' interference
experiments have been done typically at only a single energy and
typically many exciton binding energies above the band-gap.
Moreover, these initial experiments lacked an absolute calibration
of the relative optical phase, and thus were insensitive to the
intrinsic phase. Such a calibration is possible \cite{Schumacher96},
and could be used to verify the predictions we present here. A
nontrivial intrinsic phase would have implications for the use of
`$1+2$' current injection as an absolute measurement of the
carrier-envelope phase of an ultrashort optical pulse
\cite{Roos03,Fortier04,Roos05}.

In the present work, we extend the theory of `1+2' coherent control
of bulk semiconductor interband transitions beyond the independent
particle approximation, employing a set of approximations that are
valid close to the $\Gamma $ point of a direct gap bulk
semiconductor. Our investigation is limited to a perturbative
treatment in the fields. In this limit of low photoinjected carrier
density, the only inter-particle interaction of importance is that
between a single electron and hole. We show that, due to the
electron-hole attraction, a nontrivial phase shift does in fact
occur in the control of current and spin current, but not in the
control of carrier population or spin polarization. The intrinsic
phase can be understood in terms of partial wave phase shifts due to
the Coulomb potential between electron and hole. In addition, we
find an enhancement of each process, and relate it to the Coulomb
enhancements of one- and two-photon absorption.

In the next section, we establish notation necessary to describe the
`1+2' processes in terms of one- and two-photon transition
amplitudes. In section \ref{sec:model}, we present the microscopic
model. The transition amplitudes are worked out in section
\ref{sec:transition_amplitudes}. The final expressions for the `1+2'
effects are given in section \ref{sec:results}, and numerical
results for GaAs are presented. In section \ref{sec:discussion},
further understanding of the enhancement and intrinsic phase is
discussed, and we examine the ratios often used as figures of merit
for `1+2' effects. Intermediate state Coulomb enhancement is
examined in Appendix \ref{Appx_IntStateEnhancement}. For reference,
the current injection tensor is worked out for parabolic bands in
Appendix \ref{Appx_etaFree}.

\section{Preliminaries\label{sec:preliminaries}}

The rate of photocurrent injection into an unbiased bulk
semiconductor by a two color light field
$\mathbf{E}(t)=\mathbf{E}_{\omega }\exp (-i\omega t)+
\mathbf{E}_{2\omega } \exp(-i2\omega t)+c.c.$ can be written
\begin{equation}
\frac{dJ}{dt}^{i}=\eta_{(I)} ^{ijkl}E_{\omega }^{*j}E_{\omega
}^{*k}E_{2\omega }^{l}+c.c.,  \label{current phenomenological}
\end{equation}
where superscripts denote Cartesian components and repeated indices
are to be summed over \cite{Atanasov96}. The fourth rank tensor
$\eta_{(I)} $, called the current injection tensor, describes the
material response. It is purely imaginary in the independent
particle approximation \cite{Atanasov96}, but can be complex in
general. We define the intrinsic phase $\delta ^{ijkl}$ of the
component $\eta ^{ijkl}$ as
\begin{equation}
\delta ^{ijkl}=\arctan \left( \mathrm{Re} \left( \eta _{(I)}^{ijkl}
\right) /\mathrm{Im} \left( \eta _{(I)}^{ijkl}\right)\right)
\label{delta explicitly defined}
\end{equation}
so that it is zero or $\pi $ in the independent particle
approximation. When the electron-hole interaction is included in the
set of approximations used below, all the components of $\eta_{(I)}$
have the same phase. That is,
\begin{equation}
\eta_{(I)} ^{ijkl}=ie^{i\delta }\left| \eta_{(I)} ^{ijkl}\right| .
\end{equation}
The intrinsic phase $\delta $ appears as a phase shift in the
dependence of the current injection on the phase of the optical
fields. For co-linearly polarized fields ($\mathbf{E}_{\omega
}=E_{\omega }\exp (i\phi _{\omega })\mathbf{\hat{x}}$ and
$\mathbf{E}_{2\omega }=E_{2\omega }\exp (i\phi _{2\omega
})\mathbf{\hat{x}}$), for example, the current injection is
\[
\frac{d\mathbf{J}}{dt}=2E_{\omega }^{2}E_{2\omega }\left|
\eta_{(I)} ^{xxxx}\right| \sin \left( 2\phi _{\omega }-\phi
_{2\omega }-\delta \right) \mathbf{\hat{x}}.
\]
We are ignoring scattering processes through which the current would
relax to a steady state value under continuous illumination, or
would decay to zero following pulsed excitation. The current
injection discussed here can be used as a source term in
hydrodynamic equations that treat the scattering phenomenologically
\cite{Atanasov96,HacheIEEE98,Cote99} or in microscopic transport
equations \cite{Kral00}. Coulomb effects other than the excitonic
effects we consider here play a role in scattering, especially at
high densities of excited carriers. Such Coulomb effects are outside
the scope of this paper.

The current injection tensor can be written in terms of one- and
two-photon transition amplitudes. We take as the initial state a
clean, cold semiconductor. In a Fermi's golden rule calculation for
the ballistic current, light produces transitions to final states
$\left| n\right\rangle $ with velocity $\mathbf{v}_{nn}$,
probability amplitude $c_{n}$, and energy $\hbar \omega _{n}$ above
the ground state. The final states will be specified in detail in
the next section; here the label $n$ represents the set of quantum
numbers for either an interacting or independent electron-hole pair.
Thus,
\begin{eqnarray}
\frac{d}{dt}\mathbf{J} &=&\frac{e}{L^{3}}\sum_{n}
\mathbf{v}_{nn}\frac{d}{dt}
\left| c_{n}\right| ^{2}  \label{currentMicro} \\
&=&\frac{2\pi e}{L^{3}}\sum_{n}\mathbf{v}_{nn}\left| \Omega
_{n}^{(1)}+\Omega _{n}^{(2)}\right| ^{2}\delta \left( 2\omega
-\omega _{n}\right)  \nonumber \\
&=&\frac{2\pi e}{L^{3}}\sum_{n}\mathbf{v}_{nn}\left\{ \left| \Omega
_{n}^{(1)}\right| ^{2}+\left| \Omega _{n}^{(2)}\right| ^{2}+\left( \Omega
_{n}^{(1)}\Omega _{n}^{(2)*}+c.c.\right) \right\} \delta \left( 2\omega
-\omega _{n}\right) ,  \nonumber
\end{eqnarray}
where $e$ is the electron charge (negative), $L^{3}$ is a
normalization volume, and $\Omega _{n}^{\left( i\right) }$ is the
amplitude for an \textit{i}-photon transition. The $\Omega
_{n}^{\left( i\right) }$ take the form
\begin{eqnarray}
\Omega _{n}^{(1)} &=&\mathbf{E}_{2\omega }\cdot \mathbf{D}_{n}^{\left(
1\right) }  \label{D1definition}\\
\Omega _{n}^{(2)} &=&\mathbf{E}_{\omega }\mathbf{E}_{\omega
}:\mathsf{D} _{n}^{\left( 2\right) }, \label{D2definition}
\end{eqnarray}
where the vector $\mathbf{D}_{n}^{\left( 1\right) }$ and second rank
tensor $\mathsf{D}_{n}^{\left( 2\right) }$ depend only on properties
of the material. We consider them in detail in section
\ref{sec:transition_amplitudes}.

In the last equation of (\ref{currentMicro}), the first term is
(one-photon) one-color current injection, the second term describes
two-photon one-color current injection, while the interference term
describes the `1+2' current. Because of their different dependencies
on the electric field amplitudes, these three terms can in principle
be separated experimentally. But the first two terms vanish for
centrosymmetric crystals, and the first vanishes even for zincblende
crystals; we neglect them here. Excitonic effects on the first term
were studied by Shelest and \'{E}ntin \cite{Shelest79,Belinicher80}.
The third term survives in all materials. Comparing its expression
with the phenomenological form (\ref{current phenomenological}), we
have
\begin{equation}
\eta_{(I)} ^{ijkl}=\frac{2\pi e}{L^{3}}\sum_{n}v_{nn}^{i}\left(
D_{n}^{(2)*}\right) ^{jk}\left( D_{n}^{(1)}\right) ^{l}\delta
\left( 2\omega -\omega _{n}\right)  \label{etaSchematic}.
\end{equation}

Even if scattering from impurities and phonons is neglected, the
injection current described by (\ref{currentMicro}) does not capture
the full current density; there are also optical rectification and
`shift' contributions to the current \cite{Aversa96}. The one-color
varieties of these have been studied in some detail
\cite{Belinicher80,SturmanFridkin,SipeShkrebtii00,Cote02}, but the
`1+2' varieties have not. However, the different time dependencies
of the three current contributions allows for their separate
examination experimentally, at least in principle, and rough
order-of-magnitude estimates indicate that typically the injection
current will be the largest; it is the only contribution to the
current we treat here.

As in (\ref{currentMicro}), the carrier injection rate can be
written as
\begin{eqnarray}
\frac{d}{dt}n &=&\frac{1}{L^{3}}\sum_{n}\frac{d}{dt}\left|
c_{n}\right| ^{2}
\label{carrier injection rate} \\
&=&\frac{2\pi}{L^{3}}\sum_{n}\left\{ \left| \Omega
_{n}^{(1)}\right| ^{2}+\left| \Omega _{n}^{(2)}\right| ^{2}+\left(
\Omega _{n}^{(1)}\Omega _{n}^{(2)*}+c.c.\right) \right\} \delta
\left( 2\omega -\omega _{n}\right) , \nonumber
\end{eqnarray}
where $n$ is the number density of injected electron-hole pairs. The
first two terms in (\ref{carrier injection rate}) are the usual one-
and two-photon absorption rates, which we denote by
$\dot{n}_{2\omega }$ and $\dot{n}_{\omega }$ respectively. In terms
of one- and two-photon coefficients $\xi _{\left( 1\right) }^{ij}$
and $\xi _{\left( 2\right) }^{ijkl}$ that depend only on the
properties of the material, they can be written  as
$\dot{n}_{2\omega } =\xi _{\left( 1\right) }^{ij}E_{2\omega
}^{i}E_{2\omega }^{*j}$ and $\dot{n}_{\omega } =\xi _{\left(
2\right) }^{ijkl}E_{\omega }^{i}E_{\omega }^{j}E_{\omega
}^{*k}E_{\omega }^{*l}$ \cite{Atanasov96}. From (\ref{carrier
injection rate}),
\begin{eqnarray}
\xi _{\left( 1\right) }^{ij}&=&\frac{2\pi }{L^{3}}\sum_{n}\left(
D_{n}^{(1)*}\right) ^{i}\left( D_{n}^{(1)}\right) ^{j}\delta
\left( 2\omega -\omega _{n}\right) \label{one_photon_schematic}\\
\xi _{\left( 2\right) }^{ijkl}&=&\frac{2\pi }{L^{3}}\sum_{n}\left(
D_{n}^{(2)*}\right) ^{ij}\left( D_{n}^{(2)}\right) ^{kl}\delta
\left( 2\omega -\omega _{n}\right) \label{two photon schematic}.
\end{eqnarray}
The third term in
(\ref{carrier injection rate}), denoted $\dot{n}_{I}$, allows
population control as discussed and observed by Fraser \textit{et
al} \cite{Fraser99PRL, Fraser03}. It can be written in terms of a
third rank tensor $\xi _{\left( I\right) }^{ijk}$ as
\cite{Fraser99PRL}
\begin{equation}
\dot{n}_{I} = \xi _{\left( I\right) }^{ijk} E_{\omega
}^{*i}E_{\omega }^{*j}E_{2\omega}^{k} + c.c.,
\end{equation}
where
\begin{equation}
\xi_{(I)}^{ijk}=\frac{2\pi }{L^{3}}\sum_{n}\left(
D_{n}^{(2)*}\right) ^{ij}\left( D_{n}^{(1)}\right) ^{k}\delta
\left( 2\omega -\omega _{n}\right)  \label{xiSchematic}.
\end{equation}
It is purely real in the independent particle approximation
\cite{Fraser99PRL, Fraser03}.

Expressions such as (\ref{currentMicro}) and (\ref{carrier injection
rate}) can also be written for carrier spin polarization and spin
current \cite{StevensReview04}. The interference terms of these
describe `1+2' spin control \cite{Stevens_pssb} and `1+2' spin
current injection \cite{BhatSipe00}, which can be written in terms
of material response pseudotensors $\zeta^{ijkl}_{(I)}$
\cite{Stevens_pssb} and $\mu^{ijklm}_{(I)}$ \cite{Najmaie03},
respectively. In the independent particle approximation,
$\zeta^{ijkl}_{(I)}$ is purely imaginary \cite{Stevens_pssb}, while
$\mu^{ijklm}_{(I)}$ is purely real \cite{Najmaie03}.

The phases of the material response tensors $\eta_{(I)}$,
$\xi_{(I)}$, $\mu_{(I)}$, and $\zeta_{(I)}$ are related to the
phases of the one- and two-photon matrix elements,
$\mathbf{D}_{n}^{(1)}$ and $\mathsf{D}_{n}^{\left( 2\right) }$.
The one- and two-photon matrix elements also appear, respectively,
in the one- and two-photon absorption coefficients, as can be seen
from (\ref{one_photon_schematic}) and (\ref{two photon
schematic}). There have been many theoretical investigations of
one- and two-photon absorption near the direct gap of bulk
semiconductors that include excitonic effects
\cite{Dimmock67,Nathan85}. However, since one- and two-photon
absorption are insensitive to the phases of the transition
amplitudes, those calculations took no care to get the phases of
the transition amplitudes correct. In the next two sections we
find the transition amplitudes with the correct phases including
excitonic effects.

\section{Model\label{sec:model}}

We first review the two-band effective mass model of Wannier
excitons; the two bands are nondegenerate conduction and valence
bands that are parabolic and isotropic with a direct gap
$E_{cv}^{g}$ at $\mathbf{k}=\mathbf{0}$ (the $\Gamma $ point)
\cite{YuCardonaChapter6,BassaniBook}. It has been used to study
excitonic effects on one-photon absorption \cite{Elliott57},
two-photon absorption \cite{Mahan68}, and other nonlinear optical
processes \cite{Blossey70, Ganguly67, Martin71, GarciaCristobal94,
GarciaCristobal98, Kolber78, SheikBahae94}. We subsequently describe
a generalization that accounts for degeneracy and multiple bands. It
has been used for two-photon absorption \cite{LeeFan74}, and has
been implied whenever two-band results have been applied to actual
semiconductors.

The total Hamiltonian of the system can be written in the form
$H=H_{B}+H_{C}+H_{\text{int}}(t)$. Here, $H_{0}=H_{B}+H_{C}$ is the
field-free Hamiltonian made up of the single-particle part $H_{B}$
and the part due to the Coulomb interaction between carriers
$H_{C}$. Using the minimal coupling Hamiltonian, the optical
perturbation takes the form $H_{\text{int}}(t)=-\left( e/c\right)
\mathbf{A}(t)\cdot \mathbf{v}+e^{2}A^{2}/(2mc^{2})$, where
$\mathbf{A}(t)$ is the vector potential associated with the Maxwell
electric field and $\mathbf{v}$ is the velocity operator associated
with $H_{0}$. In the long wavelength limit, the position dependence
of $\mathbf{A}(t)$ is neglected, and thus the second term in
$H_{\text{int}}(t)$ may be neglected, since it can simply be
absorbed in an overall time-dependent phase of the full system ket
and hence cannot cause any transitions between states of $H_{0}$.
Many approximate approaches to band structure calculation, including
most pseudopotentials, and the truncation to a finite number of
bands, implicitly assume an underlying field-free Hamiltonian that
is nonlocal; there is then a correction to the interaction
Hamiltonian $H_{\text{int}}(t)$ in the velocity gauge
\cite{Girlanda81, Aversa95}. However, we neglect such nonlocal
corrections, as has been the practice in previous calculations of
coherent control effects \cite{Atanasov96, Fraser99PRL, BhatSipe00}.

The initial state is the ``vacuum'' $\left| 0\right\rangle $; it
corresponds to completely filled valence bands and empty conduction
bands. If the Coulomb interaction were neglected in a two-band model
consisting of valence ($v$) and conduction ($c$) bands, the final
states would be of the form $a_{c\mathbf{k}}^{\dagger
}a_{v\mathbf{k}}\left| 0\right\rangle $, where the operator
$a_{n\mathbf{k}}^{\dagger }$ creates an electron in an eigenstate of
$H_{B}$, a Bloch state $\left| n,\mathbf{k}\right\rangle $ with band
index $n$ and wavevector $\mathbf{k}$. The photon momentum has been
neglected, consistent with the long wavelength approximation. The
Coulomb interaction couples states at different $\mathbf{k}$; thus
including $H_{C}$ the final states are of the form
\begin{equation}
\left| cv\bm{\kappa}\right\rangle \equiv
\sum\limits_{\mathbf{k}}A_{cv}^{\bm{\kappa }}\left(
\mathbf{k}\right) a_{c\mathbf{k}}^{\dagger }a_{v\mathbf{k}}\left|
0\right\rangle ,  \label{final state}
\end{equation}
where $\bm{\kappa }$ labels the state; its physical meaning is given
below. In the effective mass Wannier exciton approximation, the
Fourier transform
\begin{equation}
\psi _{cv}^{\bm{\kappa
}}(\mathbf{r})=\sum\limits_{\mathbf{k}}A_{cv}^{\bm{\kappa }}\left(
\mathbf{k} \right) e^{i\mathbf{k}\cdot \mathbf{r}},
\label{FourierTransform}
\end{equation}
which is the wavefunction for the relative coordinate between
electron and hole, is a hydrogenic wavefunction satisfying
\begin{equation}
-\frac{\hbar ^{2}}{2\mu _{cv}}\nabla ^{2}\psi _{cv}^{\bm{\kappa
}}(\mathbf{r})-V(r)\psi _{cv}^{\bm{\kappa }}(\mathbf{r})=\left(
E_{cv}\left( \bm{\kappa }\right) -E_{cv}^{g}\right) \psi
_{cv}^{\bm{\kappa }}(\mathbf{r}), \label{exciton_schrodinger}
\end{equation}
where $\mu _{cv}^{-1}=m_{c}^{-1}+m_{v}^{-1}$ is the reduced mass in
terms of the (positive) conduction and valence band effective
masses, and $V\left( r\right) $ is the Coulomb potential,
$V(r)=e^{2}/\left( \epsilon r\right) $, screened by the static
dielectric constant $\epsilon $
\cite{Elliott57,Baldereschi71,HaugKoch}. The state has energy
\[
E_{cv}\left( \bm{\kappa }\right) =\frac{\hbar ^{2}\kappa
^{2}}{2\mu _{cv}}+E_{cv}^{g}.
\]
We choose the states to be normalized over the volume $L^{3}$ by
$\langle m, \mathbf{k} | n, \mathbf{k} \rangle=\delta_{n,m}
\delta_{\mathbf{k},\mathbf{k}^{\prime}}$ and $\langle cv\bm{\kappa}
| cv\bm{\kappa}^{\prime} \rangle=
\delta_{\bm{\kappa},\bm{\kappa}^{\prime}}$; as a result $\psi
_{cv}^{\bm{\kappa }}(\mathbf{r})$ is unitless, having the
normalization $\int d^{3}r(\psi _{cv}^{\bm{\kappa
}}(\mathbf{r}))^{*}\psi _{cv}^{\bm{\kappa }^{\prime
}}(\mathbf{r})=L^{3} \delta_{\bm{\kappa},\bm{\kappa}^{\prime}}$.

Our focus in this paper is on the unbound solutions to
(\ref{exciton_schrodinger}); bound exciton states lack relative
velocity between the electron and hole, and hence do not contribute
to the ballistic current or spin current. For a Fermi's golden rule
calculation of the current or spin current, the unbound state must
behave asymptotically like an outgoing plane wave in the relative
coordinate between electron and hole; $\bm{\kappa}$ is the
wavevector of the outgoing plane wave. Specifically, we must use
``ionization states'' rather than scattering states
\cite{BreitBethe54}, as was done for atomic `1+2' ionization
\cite{Baranova90}. They are related by $\left( \psi
_{cv}^{\bm{\kappa }}(\mathbf{r})\right) _{\text{ion}}=\left[ \left(
\psi _{cv}^{-\bm{\kappa }}(\mathbf{r})\right) _{\text{scatt}}\right]
^{*}$ \cite{TaylorScatteringTheory}. Calculations of one- or
two-photon absorption are insensitive to an error in this choice of
boundary condition, but the present calculation is not, since it is
sensitive to the relative phase of the transition amplitudes.

The ionization state wavefunctions that solve
(\ref{exciton_schrodinger}) can be expressed as a sum over partial
waves,
\begin{equation}
\psi _{cv}^{\bm{\kappa }}(\mathbf{r})=e^{\frac{\pi }{2a_{cv}\kappa
}} \sum_{l=0}^{\infty }\frac{\Gamma \left( l+1+\frac{i}{a_{cv}\kappa
}\right) }{\left( 2l\right) !}\left( 2i\kappa r\right)
^{l}e^{-i\kappa r}P_{l}\left( \frac{\mathbf{r}\cdot
\bm{\kappa}}{r\kappa }\right) {}_{1}F_{1}\left(
l+1+\frac{i}{a_{cv}\kappa };2l+2;2i\kappa r\right) ,
\label{ionization state}
\end{equation}
where $a_{cv}=\epsilon \hbar ^{2}/\left( \mu _{cv}e^{2}\right) $ is
the exciton Bohr radius, and $\kappa $ and $r$ mean $\left|
\bm{\kappa }\right| $ and $\left| \mathbf{r}\right|
$\cite{footnote:IonizationState}\nocite{Schiffp119}. The $P_{l}$ are
Legendre polynomials, $_{1}F_{1}$ is a confluent hypergeometric
function, and $\Gamma$ is the Gamma function.

Such a two-band model of Wannier excitons is useful for the
description of many optical properties. However, near the band gap
at the $\Gamma $ point of a typical zincblende semiconductor there
are, counting spin, eight bands: two each of conduction ($c$), heavy
hole ($hh$), light hole ($lh$) and split-off hole ($so$). Other
bands, especially the next higher conduction bands, can also be
important for some processes, especially for population and spin
control.

The existence of multiple bands and band degeneracy modifies the
exciton Hamiltonian, i.e., the operator acting on $\psi
_{cv}^{\bm{\kappa }}(\mathbf{r})$ in the left side of
(\ref{exciton_schrodinger}). In the effective mass approximation,
using a basis of $\Gamma $ point states, the kinetic part of the
Wannier exciton Hamiltonian has a matrix structure
\cite{Luttinger55,Dresselhaus56}. Even though this approximation
neglects band warping, nonparabolicity, and inversion asymmetry, the
Hamiltonian lacks analytic eigenstates \cite{Dresselhaus56}. This is
essentially due to the degeneracy of the $hh$ and $lh$ bands at the
$\Gamma$ point. As a result of the difference between $m_{hh}$ and
$m_{lh}$ there is `envelope-hole coupling,' \cite{Sondergeld77I}
which is a spin-orbit-like coupling between the orbital angular
momentum of the exciton envelope function and the total angular
momentum of the valence band $\Gamma $ point Bloch functions
\cite{Baldereschi73}. Baldereschi and Lipari split the effective
mass Hamiltonian into a sum of terms based on symmetry, and showed
that in a spherical approximation the envelope-hole coupling could
be treated as a perturbation to the diagonal part, which has
analytic, hydrogenic eigenstates \cite{Baldereschi70,Baldereschi71}.
In order to extract the main physics, while preserving the
simplicity of the two-band model, we neglect envelope-hole coupling
entirely. In this approximation, (\ref{exciton_schrodinger}) remains
valid for each conduction-valence band pair, however one must use
`average' effective masses for degenerate bands. Specifically, the
effective mass of the valence bands $hh$, $lh$, and $so$ is $m/
\gamma_{1L}$, where $m$ is the free electron mass, and $\gamma_{1L}$
is one of the Luttinger parameters \cite{Baldereschi71}. The upper
conduction bands have a different average effective mass. Note that
$\psi _{cv}^{\bm{\kappa }}(\mathbf{r})$ is independent of $c$ and
$v$ within the set of bands $\left\{ c,hh,lh,so \right\}$. The
effect of envelope-hole coupling has been studied for exciton bound
states \cite{Baldereschi70,Baldereschi71,Sondergeld77II}, but not
for optical processes involving unbound excitons in the continuum.

Even within this model, the presence of multiple bands provides two
types of terms in the sum over intermediate states in the two-photon
amplitude: two-band terms, in which the intermediate and final
states are in the same exciton series [i.e., two states of the form
(\ref{final state}) with the same $c$ and $v$], and three-band
terms, in which the intermediate and final states are in different
series. Three-band terms are important for some processes but not
for others. For current control, three-band terms are important for
cross-linearly polarized fields \cite{BhatSipe_unpub}, and for spin
current control, they are important for the spin current due to
co-linearly polarized fields \cite{BhatSipe00}. Three-band terms are
essential for population and spin control
\cite{StevensReview04,BhatSipe_unpub}.

The velocity matrix elements involving the state $\left|
cv\bm{\kappa } \right\rangle $ are
\begin{equation}
\left\langle cv\bm{\kappa }\right| \mathbf{v}\left| 0\right\rangle
=\sum_{\mathbf{k}}\left( A_{cv}^{\bm{\kappa }}\left(
\mathbf{k}\right) \right) ^{*}\mathbf{v}_{cv}\left(
\mathbf{k}\right)  \label{vac_pairVelocity}
\end{equation}
\begin{equation}
\left\langle cv\bm{\kappa }\right| \mathbf{v}\left| c^{\prime
}v^{\prime }\bm{\kappa }^{\prime }\right\rangle
=\sum\limits_{\mathbf{k}}\left( A_{cv}^{\bm{\kappa }}\left(
\mathbf{k}\right) \right) ^{*}A_{c^{\prime }v^{\prime
}}^{\bm{\kappa }^{\prime }}\left( \mathbf{k}\right) \left(
\mathbf{v}_{cc^{\prime }}\left( \mathbf{k}\right) \delta
_{v,v^{\prime }}-\mathbf{v}_{v^{\prime }v}\left( \mathbf{k}\right)
\delta _{c,c^{\prime }}\right) , \label{twotwovelocity}
\end{equation}
where $\mathbf{v}_{nm}\left( \mathbf{k}\right) \equiv \left\langle
n,\mathbf{k}\right| \mathbf{v}\left| m,\mathbf{k}\right\rangle $ is
the velocity matrix element between Bloch states.

\section{Transition amplitudes\label{sec:transition_amplitudes}}

In the independent particle approximation, the transition amplitudes
are
\begin{equation}
\Omega _{cv\bm{\kappa }}^{(1\text{-free})}=i\frac{e}{2\hbar \omega
}\mathbf{E}_{2\omega }\cdot \mathbf{v}_{cv}\left( \bm{\kappa}
\right) , \label{one_photon_free}
\end{equation}
and $\Omega _{cv\bm{\kappa }}^{(2\text{-free})}=
\sum_{c^{\prime},v^{\prime}} \Omega _{cc^{\prime }vv^{\prime
}\bm{\kappa }}^{(2\text{-free})}$, where
\begin{equation}
\Omega _{cc^{\prime }vv^{\prime }\bm{\kappa }}^{(2\text{-free})}
\equiv \left( \frac{e}{\hbar \omega }\right) ^{2}\frac{\left(
\mathbf{E}_{\omega }\cdot \left( \mathbf{v}_{cc^{\prime }}\left(
\bm{\kappa }\right) \delta _{v,v^{\prime }}-\delta _{c,c^{\prime
}}\mathbf{v}_{v^{\prime }v}\left( \bm{\kappa }\right) \right)
\right) \left( \mathbf{E}_{\omega }\cdot
\mathbf{v}_{c^{\prime}v^{\prime }}\left( \bm{\kappa }\right) \right)
}{E_{c^{\prime }v^{\prime }}/\hbar-\omega\left( \bm{\kappa}\right)}
\label{two photon_free}.
\end{equation}

With excitonic effects included, using the perturbation
$H_{\text{int}}\left( t\right) $ to second order gives the
transition amplitudes
\begin{equation}
\Omega _{cv\bm{\kappa }}^{(1)}=\frac{ie}{2\hbar \omega
}\mathbf{E}_{2\omega }\cdot \left\langle cv\bm{\kappa }\right|
\mathbf{v}\left| 0\right\rangle \label{one photon rate},
\end{equation}
and
\begin{equation}
\Omega _{cv\bm{\kappa }}^{(2)}=\left( \frac{e}{\hbar \omega
}\right) ^{2}\sum_{c^{\prime }v^{\prime }\bm{\kappa }^{\prime
}}\frac{\left( \mathbf{E}_{\omega }\cdot \left\langle cv\bm{\kappa
}\right| \mathbf{v}\left| c^{\prime }v^{\prime }\bm{\kappa
}^{\prime }\right\rangle \right) \left( \mathbf{E}_{\omega }\cdot
\left\langle c^{\prime }v^{\prime }\bm{\kappa }^{\prime }\right|
\mathbf{v}\left| 0\right\rangle \right) }{E_{c^{\prime }v^{\prime
}}\left( \bm{\kappa }^{\prime }\right) /\hbar -\omega },
\label{two photon rate}
\end{equation}
where the sum over intermediate states is over both bound and free
excitons. The two-photon transition amplitude is more difficult to
deal with due to the sum over intermediate states; however, in our
set of approximations it has been done exactly
\cite{Mahan68,LeeFan74,Rustagi73}.

In order to proceed analytically, it is common to use
(\ref{vac_pairVelocity}) and (\ref{twotwovelocity}), and then make
an expansion in $\mathbf{k}$ of the velocity matrix elements
$\mathbf{v}_{nm}\left( \mathbf{k}\right) $ about the $\Gamma $ point
\cite{Elliott57, Loudon62, Mahan68, Rustagi73, LeeFan74, HaugKoch}.
However, due to the degeneracy at the $\Gamma $ point, the
coefficients of such an expansion can depend on the direction of
$\mathbf{k}$ \cite{BirPikus}. To proceed, we note that Wannier
excitons have large spatial extent and hence only a small region of
wavevectors is important for them, i.e., $A_{cv}^{\bm{\kappa
}}\left( \mathbf{k}\right)$ is peaked in the region of $\mathbf{k}$
near $\bm{\kappa}$ \cite{BassaniBook}. This is especially true for
final states with energies above the band gap. Thus, we expand
$\mathbf{v}_{nm}\left( \mathbf{k}\right) $ about the $\Gamma$ point,
approached in the direction $\mathbf{\hat{\bm{\kappa}}}$,
\begin{equation}
v_{nm}^{i}\left( \mathbf{k}\right) =v_{nm}^{i}\left(
\mathbf{\hat{\bm{\kappa}}} \right) +\mathbf{k}\cdot
\bm{\nabla}_{\mathbf{k}}v_{nm}^{i}(\mathbf{\hat{\bm{\kappa}}}
)+\ldots , \label{Bloch_v_expansion}
\end{equation}
where $v_{nm}^{i}\left( \mathbf{\hat{\bm{\kappa}}} \right) \equiv
\lim_{\lambda \rightarrow 0} \left\langle n,\Gamma \right|
\mathbf{v}\left| m, \lambda \bm{\kappa} \right\rangle$ and
$\bm{\nabla}_{\mathbf{k}} v_{nm}^{i} \left(
\mathbf{\hat{\bm{\kappa}}} \right) \equiv \lim_{\lambda \rightarrow
0} \bm{\nabla}_{\bm{\kappa}} \left\langle n,\Gamma \right|
\mathbf{v}\left| m, \lambda \bm{\kappa} \right\rangle$.

Optical transitions due to the first term in
(\ref{Bloch_v_expansion}) are called `allowed', while those due to
the second term are called `forbidden'. We restrict ourselves to
materials for which the `allowed' valence to conduction band
transition does not vanish. Keeping only the `allowed' term in
(\ref{vac_pairVelocity}),\cite{Elliott57}
\begin{equation}
\left\langle cv\bm{\kappa }\right| \mathbf{v}\left| 0\right\rangle
=\mathbf{v}_{cv}\left( \mathbf{\hat{\bm{\kappa}}} \right) \ \left(
\psi _{cv}^{\bm{\kappa }}(\mathbf{r}=\mathbf{0})\right) ^{*}.
\label{allowed_interband_v}
\end{equation}
For the intravalence and intraconduction band transitions, the first
two terms of (\ref{Bloch_v_expansion}) in (\ref{twotwovelocity})
give \cite{Rustagi73}
\begin{equation}\label{v_intra}
\begin{split}
\left\langle cv\bm{\kappa }\right| \mathbf{v}\left| c^{\prime}
v^{\prime }\bm{\kappa }^{\prime }\right\rangle =&\left[ \delta
_{v,v^{\prime }} \mathbf{v}_{cc^{\prime }}\left(
\mathbf{\hat{\bm{\kappa}}} \right) -\delta _{c,c^{\prime }}
\mathbf{v}_{v^{\prime }v}\left( \mathbf{\hat{\bm{\kappa}}} \right)
\right] \int \frac{d^{3}r}{L^{3}}\left( \psi _{cv}^{\bm{\kappa
}}\left( \mathbf{r}\right) \right) ^{*}\psi
_{c^{\prime }v^{\prime }}^{\bm{\kappa }^{\prime }}(\mathbf{r}) \\
&-\left[ \delta _{v,v^{\prime
}}\bm{\nabla}_{\mathbf{k}}v_{cc^{\prime }}^{i}\left(
\mathbf{\hat{\bm{\kappa}}} \right) -\delta _{c,c^{\prime
}}\bm{\nabla }_{\mathbf{k}}v_{v^{\prime }v}^{i}\left(
\mathbf{\hat{\bm{\kappa}}} \right) \right] i\int
\frac{d^{3}r}{L^{3}}\left( \psi _{cv}^{\bm{\kappa }}\left(
\mathbf{r}\right) \right) ^{*}\bm{\nabla }\psi _{c^{\prime
}v^{\prime }}^{\bm{\kappa }^{\prime }}(\mathbf{r}).
\end{split}
\end{equation}
In particular \cite{Rustagi73, BetheSalpeter},
\begin{equation}
\left\langle cv\bm{\kappa }\right| \mathbf{v}
\left|cv\bm{\kappa}\right\rangle = \sum_{\mathbf{k}}\left(
A_{cv}^{\mathbf{\kappa }}\left( \mathbf{k}\right) \right)
^{*}A_{cv}^{\mathbf{\kappa }}\left( \mathbf{k}\right) \frac{\hbar
}{\mu _{cv}}\mathbf{k} = -i\frac{\hbar }{\mu _{cv}}\int
\frac{d^{3}r}{L^{3}}\left( \psi _{cv}^{\mathbf{\kappa
}}(\mathbf{r})\right) ^{*}\mathbf{\nabla }_{\mathbf{r}}\psi
_{cv}^{\mathbf{\kappa }}(\mathbf{r}) = \hbar \bm{\kappa} / \mu_{cv}
\label{asymptotic_v}.
\end{equation}

For Ge and for simple models of zincblende semiconductors that
neglect lack of inversion symmetry, the first term in
(\ref{v_intra}) always vanishes. This means that there are only
`allowed-forbidden' two-photon transitions (the interband transition
is allowed, while the intraband transition is forbidden). When the
first term is nonvanishing, there are `allowed-allowed' two-photon
transitions. In principle, for materials that lack a center of
inversion, there is also a small contribution to the
`allowed-forbidden' two-photon transition from the first term in
(\ref{v_intra}) and the term in $\left\langle cv\bm{\kappa }\right|
\mathbf{v}\left| 0\right\rangle $ that comes from the second term in
(\ref{Bloch_v_expansion}); we neglect it in what follows, but note
that when compared to the dominant `allowed-forbidden' contribution
that we consider here, it has a different Coulomb enhancement but
the same intrinsic phase (see Eq.\ 2.32 of Rustagi \textit{et al}.\
\cite{Rustagi73}).

We write $\Omega _{cv\bm{\kappa }}^{(2)}= \Omega _{cv\bm{\kappa
}}^{(2\textrm{:a-f})}+ \Omega _{cv\bm{\kappa }}^{(2\textrm{:a-a})}$,
and discuss the `allowed-forbidden' and `allowed-allowed'
transitions separately.

Using (\ref{ionization state}), (\ref{allowed_interband_v}), and
(\ref{one photon rate}), the one-photon transition amplitude is
\cite{Elliott57}
\begin{eqnarray}
\Omega _{cv\bm{\kappa }}^{(1)} &=&\Omega _{cv\bm{\kappa
}}^{(1\text{-free})}\exp \left( \frac{\pi }{2a_{cv}\kappa }\right)
\Gamma \left( 1-\frac{i}{a_{cv}\kappa }\right),   \label{one photon
allowed}
\end{eqnarray}
where only the `allowed' term is kept in $\Omega _{cv\bm{\kappa
}}^{(1\text{-free})}$. The transition is to an $s$-wave. The
one-photon absorption coefficient is proportional to the norm of
$\Omega _{cv\bm{\kappa }}^{(1)}$ [see (\ref{carrier injection rate})
and (\ref{one_photon_schematic})])\cite{Elliott57}.

For `allowed-forbidden' two-photon transitions, substituting
(\ref{allowed_interband_v}) and the second term of (\ref{v_intra})
into (\ref{two photon rate}),
\begin{equation}
\Omega _{cv\bm{\kappa }}^{(2\textrm{:a-f})}=\frac{e^{2}}{\omega
^{2}\hbar } \sum_{c^{\prime }v^{\prime }} \left(
\mathbf{E}_{\omega }\cdot \mathbf{v}_{c^{\prime}v^{\prime}}\left(
\mathbf{\hat{\bm{\kappa}}} \right) \right) E_{\omega }^{i}\left(
\delta _{v,v^{\prime }}
\bm{\nabla}_{\mathbf{k}}v_{cc^{\prime}}^{i}\left(
\mathbf{\hat{\bm{\kappa}}} \right) -\delta _{c,c^{\prime
}}\bm{\nabla }_{\mathbf{k}}v_{v^{\prime }v}^{i}\left(
\mathbf{\hat{\bm{\kappa}}} \right) \right) \cdot
\mathbf{M}_{cc^{\prime }vv^{\prime }}\left( \bm{\kappa }\right) ,
\label{two_photon_amp_midway}
\end{equation}
where
\begin{equation}
\mathbf{M}_{cc^{\prime }vv^{\prime }}\left( \bm{\kappa }\right)
\equiv -i\int d^{3}r\left( \psi _{cv}^{\bm{\kappa }}\left(
\mathbf{r}\right) \right) ^{*}\bm{\nabla }G_{c^{\prime }v^{\prime
}}\left( \mathbf{r},\mathbf{0};\hbar \omega -E_{c^{\prime
}v^{\prime }}^{g}\right) ,  \label{W_defined}
\end{equation}
and we have used the Coulomb Green function,
\[
G_{cv}\left( \mathbf{r},\mathbf{r}^{\prime };\Omega \right)
=\frac{1}{L^{3}}\sum_{\bm{\kappa }}\frac{\psi _{cv}^{\bm{\kappa
}}(\mathbf{r})\left( \psi _{cv}^{\bm{\kappa }}(\mathbf{r}^{\prime
})\right) ^{*}}{E_{cv}\left( \kappa \right) -E_{cv}^{g}-\Omega },
\]
which is known analytically \cite{Mahan68}. In particular,
\begin{equation}
G_{cv}(\mathbf{r},\mathbf{0};\hbar \omega -E_{cv}^{g})=\frac{\mu
_{cv}}{2\pi r\hbar ^{2}}\Gamma \left( 1-\gamma _{cv}\right)
W_{\gamma _{cv},\frac{1}{2}}\left( \frac{2r}{a_{cv}\gamma
_{cv}}\right) ,\label{CoulombGreenFunction}
\end{equation}
where we define
\[
\gamma _{cv}\equiv \sqrt{\frac{B_{cv}}{E_{cv}^{g}-\hbar \omega }},
\]
$B_{cv}=\hbar ^{2}/\left( 2\mu _{cv}a_{cv}^{2}\right) $ is the
exciton binding energy, and $W_{\gamma ,1/2}\left( z\right) $ is a
Whittaker function with the integral representation
\begin{equation}
W_{\gamma ,1/2}\left( z\right) =\frac{ze^{-\frac{z}{2}}}{\Gamma
\left( 1-\gamma \right) }\int_{0}^{\infty }dt\left(
\frac{1+t}{t}\right) ^{\gamma }e^{-zt}. \label{Whittaker}
\end{equation}
Since the Green function depends only on the magnitude of
$\mathbf{r}$, only the \textit{p}-wave of the final state survives
the integral in Eq.\ (\ref{W_defined}) over the angles of
$\mathbf{r}$, $\int d\Omega P_{1}\left( \mathbf{\hat{r}}\cdot
\mathbf{\hat{\bm{\kappa}}}\right) \mathbf{\hat{r}}=4\pi
\mathbf{\hat{\bm{\kappa}}}/3$. The integral over $r$ can be done
using \cite{Mahan68}
\begin{equation}
\int_{0}^{\infty }r^{\sigma -1}e^{-pr}{}_{1}F_{1}\left( \alpha
;\sigma ;\lambda r\right) dr=\Gamma \left( \sigma \right)
\frac{p^{\alpha -\sigma }}{\left( p-\lambda \right) ^{\alpha }}.
\label{integral identity}
\end{equation}
The final result is
\begin{equation}
\Omega _{cv\bm{\kappa }}^{(2\textrm{:a-f})}=\exp \left( \frac{\pi
}{2a_{cv}\kappa }\right) \Gamma \left( 2-\frac{i}{a_{cv}\kappa
}\right) \sum_{c^{\prime }v^{\prime
}}N^{(\textrm{a-f})}_{cc^{\prime }vv^{\prime }}\left( \kappa
\right) \Omega _{cc^{\prime }vv^{\prime }\bm{\kappa
}}^{(2\text{-free})}, \label{two photon a-f}
\end{equation}
where only the `allowed-forbidden' term is kept in $\Omega
_{cc^{\prime }vv^{\prime }\bm{\kappa }}^{(2\text{-free})}$, and
\begin{equation}
N^{(\textrm{a-f})}_{cc^{\prime }vv^{\prime }}\left( \kappa \right)
\equiv \left( 1+a_{c^{\prime }v^{\prime }}^{2}\kappa ^{2}\gamma
_{c^{\prime }v^{\prime }}^{2}\right) 2\int_{0}^{1}\frac{S\left(
\frac{1+S}{1-S}\right) ^{\gamma _{c^{\prime }v^{\prime }}}\exp
\left( -\frac{2}{a_{cv}\kappa }\arctan \left( a_{c^{\prime
}v^{\prime }}\kappa \gamma _{c^{\prime }v^{\prime }}S\right) \right)
}{\left( 1+a_{c^{\prime }v^{\prime }}^{2}\kappa ^{2}\gamma
_{c^{\prime }v^{\prime }}^{2}S^{2}\right) ^{2}}dS. \label{N_af}
\end{equation}

For `allowed-allowed' two-photon transitions, substituting
(\ref{allowed_interband_v}) and the first term of (\ref{v_intra})
into (\ref{two photon rate}),
\begin{equation*}
\begin{split}
\Omega _{cv\bm{\kappa }}^{(2\textrm{:a-a})}=& \left(
\frac{e}{\hbar \omega }\right) ^{2}\sum_{c^{\prime }v^{\prime
}}\mathbf{E}_{\omega }\cdot \left[ \delta _{vv^{\prime
}}v_{cc^{\prime }}^{i}\left( \mathbf{\hat{\bm{\kappa}}} \right)
-\delta _{c,c^{\prime }}v_{v^{\prime }v}^{i}\left(
\mathbf{\hat{\bm{\kappa}}} \right) \right] \mathbf{E}_{\omega
}\cdot \mathbf{v}_{c^{\prime }v^{\prime }}\left(
\mathbf{\hat{\bm{\kappa}}} \right) \\ &\times \hbar \int
d^{3}r\left( \psi _{cv}^{\bm{\kappa }}\left( \mathbf{r}\right)
\right) ^{*}G_{c^{\prime }v^{\prime }}\left(
\mathbf{r},\mathbf{0};\hbar \omega -E_{c^{\prime }v^{\prime
}}^{g}\right) .
\end{split}
\end{equation*}
Since $G_{c^{\prime }v^{\prime }}$ depends only on the magnitude of
$\mathbf{r}$ [Eq. (\ref{CoulombGreenFunction})], only the $s$ part
of the final state will survive the integration over angles of
$\mathbf{r}$. Again we use (\ref{Whittaker}) for the Whittaker
function. The integral over the magnitude of $\mathbf{r}$ can be
done using an identity obtained by taking a derivative with respect
to $p$ of both sides of (\ref{integral identity}). Finally,
\begin{equation}
\Omega _{cv\bm{\kappa }}^{(2\textrm{:a-a})}= \exp \left( \frac{\pi
}{2a_{cv}\kappa }\right) \Gamma \left( 1-\frac{i}{a_{cv}\kappa
}\right) \sum_{c^{\prime }v^{\prime }}\Omega _{cc^{\prime
}vv^{\prime }\bm{\kappa }}^{(2\text{-free})}
N^{(\textrm{a-a})}_{cc^{\prime }vv^{\prime }}\left( \kappa \right),
\label{two photon a-a final}
\end{equation}
where only the `allowed-allowed' term is kept in $\Omega
_{cc^{\prime }vv^{\prime }\bm{\kappa }}^{(2\text{-free})}$, and
\begin{equation}
\begin{split}
N^{(\textrm{a-a})}_{cc^{\prime }vv^{\prime }}\left( \kappa \right)
\equiv& \left( 1+\left( a_{c^{\prime }v^{\prime }}\kappa \gamma
_{c^{\prime
}v^{\prime }}\right) ^{2}\right) 2 \\
&\times \int_{0}^{1}S\left( 1-S\frac{a_{c^{\prime }v^{\prime
}}\gamma _{c^{\prime }v^{\prime }}}{a_{cv}}\right) \left(
\frac{1+S}{1-S}\right) ^{\gamma _{c^{\prime }v^{\prime }}}\frac{\exp
\left( -\frac{2}{a_{cv}\kappa }\arctan \left( a_{c^{\prime
}v^{\prime }}\kappa \gamma _{c^{\prime }v^{\prime }}S\right) \right)
}{\left( 1+\left( a_{c^{\prime }v^{\prime }}\kappa \gamma
_{c^{\prime }v^{\prime }}S\right) ^{2}\right) ^{2}}dS. \label{N_aa}
\end{split}
\end{equation}
This agrees with Eq.\ 2.28 of Rustagi \cite{Rustagi73}, but note
that we have defined $N^{(\textrm{a-a})}_{cc^{\prime }vv^{\prime
}}\left( \kappa \right) =\left( 1+\left( a_{c^{\prime }v^{\prime
}}\kappa \gamma _{c^{\prime }v^{\prime }}\right) ^{2}\right)
I_{s,k}\left( \kappa \right) $, where $I_{s,k}\left( \kappa \right)
$ is given, with a typographical error, in Eq.\ 2.25 of that paper.

The factors $N^{(\textrm{a-f})}_{cc^{\prime }vv^{\prime }}$ and
$N^{(\textrm{a-a})}_{cc^{\prime }vv^{\prime }}$, which appear in
(\ref{N_af}) and (\ref{N_aa}) are the enhancements due to the
Coulomb interaction in the intermediate states; they are discussed
further in Appendix \ref{Appx_IntStateEnhancement}.

\section{Results\label{sec:results}}

The one- and two-photon transition amplitudes were presented in the
previous section on the basis of an expansion in $\mathbf{k}$ of the
Bloch state velocity matrix elements. The `allowed' one-photon
transition amplitude $\Omega^{(1)}$ is in (\ref{one photon
allowed}), the `allowed-forbidden' two-photon transition amplitude
is in (\ref{two photon a-f}), and the `allowed-allowed' two-photon
transition amplitude is in (\ref{two photon a-a final}). From them,
$\mathbf{D}_{cv\bm{\kappa }}^{\left( 1\right) }$ and
$\mathsf{D}_{cv\bm{\kappa }}^{\left( 2\right) }$ may be extracted by
comparison with the definitions in (\ref{D1definition}) and
(\ref{D2definition}).

\subsection{Current injection}

The `1+2' current injection is dominated by interference of
`allowed' one-photon transitions and `allowed-forbidden' two-photon
transitions \cite{BhatSipe_unpub}. Substituting these into
(\ref{etaSchematic}), and using the Gamma function identities
$\Gamma \left( x+1\right) =x\Gamma \left( x\right) $ and
\begin{equation}
\Gamma \left( 1-ix\right) \Gamma \left( 1+ix\right) =\pi x/\sinh
\left( \pi x\right),\label{Gamma identity}
\end{equation}
yields our final result for the current injection tensor
\begin{equation}
\eta_{(I)} ^{ijkl}=\sum_{c,v}\left( 1+\frac{i}{a_{cv}\kappa
_{cv}}\right) \Xi \left( a_{cv}\kappa_{cv}
\right)\sum_{c^{\prime},v^{\prime }} N^{(\textrm{a-f})}
_{cc^{\prime}vv^{\prime }} \left( \kappa _{cv}\right) \eta
_{cc^{\prime}vv^{\prime }}^{ijkl}, \label{eta_result}
\end{equation}
where
\begin{equation} \kappa_{cv}
\equiv \frac{1}{a_{cv}}\sqrt{\frac{2\hbar \omega
-E_{cv}^{g}}{B_{cv}}}, \label{energy conservation}
\end{equation}
\begin{equation}
\Xi\left( x \right)\equiv \frac{\left( \pi / x\right)\exp \left( \pi
/ x\right) }{\sinh \left( \pi / x\right)} =\frac{2\pi }{x}\left(
1-\exp \left( -2\pi /x\right) \right) ^{-1}\label{XiDefined},
\end{equation}
and
\begin{equation}
\eta_{cc^{\prime}vv^{\prime }}^{ijkl}\equiv \frac{2\pi
e}{L^{3}}\sum_{\bm{\kappa }}\Delta _{cv}^{i}\left( \bm{\kappa
}\right) \left( D_{cc^{\prime}vv^{\prime }\bm{\kappa
}}^{(2\text{-free})*}\right) ^{jk}\left( D_{cv\bm{\kappa
}}^{(1\text{-free})}\right) ^{l}\delta \left( 2\omega
-E_{cv}\left( \kappa \right) /\hbar \right) ,\label{eta_ccvv}
\end{equation}
with $\mathbf{\Delta }_{cv}\left( \bm{\kappa}\right) \equiv \left(
\mathbf{v}_{cc}\left( \bm{\kappa}\right) -\mathbf{v}_{vv}\left(
\bm{\kappa}\right) \right) =\hbar \bm{\kappa} / \mu_{cv}$ and
\begin{equation}
\left( D_{cc^{\prime}vv^{\prime }\bm{\kappa
}}^{(2\text{-free})}\right) ^{jk}=\left( \frac{e}{\hbar \omega
}\right) ^{2}\frac{\left\{ \left( \mathbf{v}_{cc^{\prime}}\left(
\bm{\kappa }\right) \delta _{v,v^{\prime }}-\delta _{c,c^{\prime
}}\mathbf{v}_{v^{\prime }v}\left( \bm{\kappa }\right) \right)
,\mathbf{v}_{c^{\prime}v^{\prime }}\left( \bm{\kappa }\right)
\right\} ^{jk}}{E_{c^{\prime}v^{\prime }}\left( \kappa \right)
/\hbar -\omega },\label{D2free}
\end{equation}
where $\left\{ \mathbf{v}_{1},\mathbf{v}_{2}\right\} ^{ij}\equiv
(v_{1}^{i}v_{2}^{j}+v_{1}^{j}v_{2}^{i})/2$ and
\begin{equation}
\left( D_{cv\bm{\kappa }}^{(1\text{-free})}\right)
^{l}=i\frac{e}{2\hbar \omega }v_{cv}^{l}\left( \bm{\kappa}\right)
.\label{D1free}
\end{equation}
Note that only the `allowed' part of (\ref{D1free}) and the
`allowed-forbidden' part of (\ref{D2free}) should be retained for a
consistent solution. We have written (\ref{eta_result}) to separate
the parts due to the electron-hole interaction. In the independent
particle approximation, the current injection tensor $\eta
_{(I\text{-free})}^{ijkl}$ is \cite{Atanasov96}
\begin{equation}
\eta _{(I\text{-free})}^{ijkl}=\sum_{c,c^{\prime},v,v^{\prime }}\eta
_{cc^{\prime}vv^{\prime }}^{ijkl};\label{eta_result_free}
\end{equation}
it is evaluated for parabolic bands in Appendix \ref{Appx_etaFree}.

\begin{figure}
\includegraphics[width=3.2in]{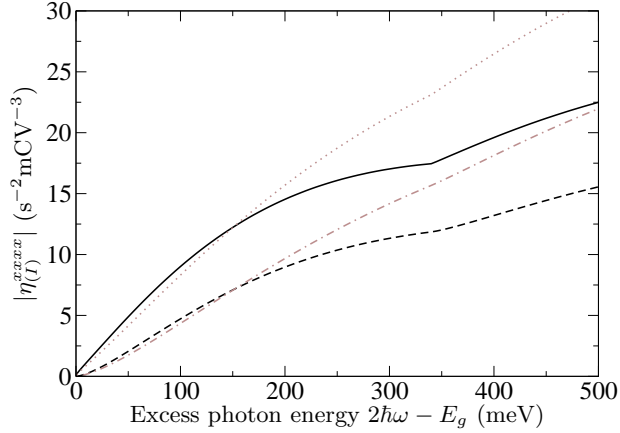}
\caption{Magnitude of the diagonal element of the current injection
tensor for GaAs with [Eq.\ (\ref{eta_result})] and without [Eq.\
(\ref{eta_result_free})] excitonic effects. The grey dotted and
dash-dotted lines are based on a parabolic band calculation of $\eta
^{xxxx}_{cc^{\prime}vv^{\prime}}$ that only includes two-band terms;
the dotted line includes excitonic effects, while the dash-dotted
line does not. The black solid and dashed lines are based on $\eta
^{xxxx}_{cc^{\prime}vv^{\prime}}$ calculated with a nonperturbative
solution of the $8\times 8$ $\mathbf{k}\cdot\mathbf{p}$ Hamiltonian;
the solid line includes excitonic effects, while the dashed line
does not. \label{Fig_eta}}
\end{figure}
For GaAs, we present in Fig.\ \ref{Fig_eta} the magnitude of
$\eta_{(I)}^{xxxx}$, based on $\eta
^{xxxx}_{cc^{\prime}vv^{\prime}}$ calculated by two methods. The
first method, described in Appendix \ref{Appx_etaFree}, uses
isotropic parabolic bands and includes only two-band terms; it uses
effective mass ratios for conduction, heavy hole, light hole, and
split-off bands of $0.067$, $0.51$, $0.082$, and $0.154$
respectively, $E_{P}=27.86$\,eV, the fundamental band gap $E_{g}$ is
1.519\,eV, and valence band spin-orbit splitting is 0.341\,eV
\cite{MadelungDataBook,PZ90}. The second method solves the $8\times
8$ $\mathbf{k}\cdot\mathbf{p}$ Hamiltonian including remote band
effects, but in a spherical approximation with warping and
spin-splitting neglected by replacing $\gamma_{2}$ and $\gamma_{3}$
with $\tilde{\gamma} \equiv \left(2\gamma_{2}+3\gamma_{3}\right)/5$
\cite{Baldereschi73}; the calculation is nonperturbative in
$\mathbf{k}$ (hence it includes band nonparabolicity) and it
includes both two- and three-band terms in the two-photon amplitude
\cite{BhatSipe_unpub}. The solid and dotted lines in Fig.\
\ref{Fig_eta} are calculated with (\ref{eta_result}), and hence
include excitonic effects; the Coulomb enhancement part of the
calculation uses $B_{cv}=4.2$\,meV \cite{Sell72} and the band
parameters listed above. Note that the solid black line in Fig.\
\ref{Fig_eta} is inconsistent in the sense that the Coulomb
enhancement is based on an expansion in $\mathbf{k}$, whereas the
free-particle result that it enhances is nonperturbative in
$\mathbf{k}$; nevertheless, such an approach has given good
agreement with experiments for one- and two-photon absorption
\cite{Sturge62,Weiler81}.

The Coulomb enhancement of $\eta_{(I)}$ can clearly be seen in Fig.\
\ref{Fig_eta}. There is a kink in each curve at excess photon energy
341 meV corresponding to the onset of transitions from the $so$
band. At higher energies, the Coulomb enhancement of $so$
transitions is larger than the Coulomb enhancements of $hh$ and $lh$
transitions, since the former transitions are to conduction band
states with lower energy. Hence, the kink in $\eta_{(I)}$ is
enhanced by excitonic effects.

\begin{figure}
\includegraphics[width=3.2in]{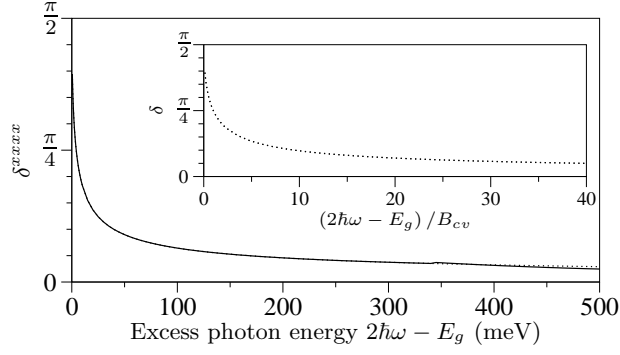}
\caption{Phase shift of the current (intrinsic phase of $\eta_{(I)}
^{xxxx}$) in GaAs due to excitonic effects. The solid line is
calculated with Eqs.\ (\ref{delta explicitly defined}) and
(\ref{eta_result}), and the dotted line is calculated with Eq.\
(\ref{phase shift}). The inset is Eq.\ (\ref{phase shift}) plotted
in scaled units.} \label{Fig_delta}
\end{figure}
We extract the intrinsic phase of $\eta_{(I)}^{xxxx}$ using
(\ref{delta explicitly defined}). The solid line in Fig.\
\ref{Fig_delta} is the intrinsic phase of $\eta_{(I)} ^{xxxx}$
calculated for GaAs with the nonperturbative $8\times 8$
$\mathbf{k}\cdot\mathbf{p}$ Hamiltonian band model; the result for
the parabolic band model is nearly identical. Since we have used a
spherical exciton model, the intrinsic phase is the same for all
components of $\eta_{(I)}^{ijkl}$. The intrinsic phase has its
maximum value of $\pi /2$ at the band edge, and goes to zero as the
light frequency increases. The decrease is smooth except for a small
kink at the onset of transitions from the $so$ band. In fact, for
excess photon energies less than the split-off energy, the intrinsic
phase has the simple analytic form
\begin{equation}
\delta \left( \omega \right) =\arctan \left(
\sqrt{\frac{B_{cv}}{2\hbar \omega -E_{g}}} \right). \label{phase
shift}
\end{equation}
Equation (\ref{phase shift}) is plotted as the dotted line in Fig.\
\ref{Fig_delta}; compared to the solid line, it is identical below
the onset of $so$ transitions, and it makes a good approximation
above the the onset of $so$ transitions. Since (\ref{phase shift})
only depends on the excess photon energy scaled by the exciton
binding energy, we plot it as a function of this scaled energy in
the inset of Fig.\ \ref{Fig_delta}; it is useful for finding the
intrinsic phase of materials other than GaAs.

In $\eta_{(I)} $ [Eq.\ (\ref{eta_result})], the two- and three-band
terms have different intermediate state Coulomb enhancement
$N^{(\textrm{a-f})}_{cc^{\prime}vv^{\prime }}$. For many materials,
however, $N^{(\textrm{a-f})}_{cc^{\prime}vv^{\prime }}$ is
approximately equal for all the terms $\eta_{cc^{\prime}vv^{\prime
}}^{ijkl}$ that contribute significantly to the total $\eta
_{(I\text{-free})}^{ijkl}$, as shown in Appendix
\ref{Appx_IntStateEnhancement} for GaAs. Thus, at photon energies
for which transitions from the heavy- and light-hole bands dominate
$\eta$, the Coulomb enhancement becomes approximately independent of
the sum over bands and we can make the simplification
\begin{equation}
\eta_{(I)} ^{ijkl}\approx F^{(I)}_{\text{a-f}} \exp \left( i\delta
\right) \eta _{(I\text{-free})}^{ijkl} \label{eta_C_simple},
\end{equation}
where the intrinsic phase is given by (\ref{phase shift}), and
\begin{equation}
F^{(I)}_{\text{a-f}}\left( \omega \right) \equiv \Xi \left(
a_{cv}\kappa_{cv} \right)\sqrt{1+\left( a_{cv}\kappa _{cv}\right)
^{-2}} N^{(\textrm{a-f})}_{ccvv}\left( \kappa _{cv}\right),
\label{enhancement}
\end{equation}
\begin{figure}
\includegraphics[width=3.2in]{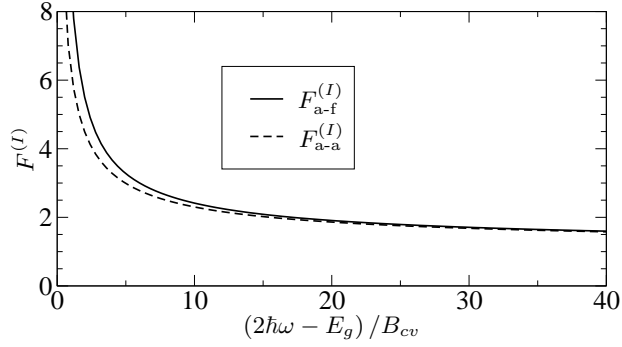}
\caption{Approximate Coulomb enhancement factors. The solid line,
applicable to current and spin current control, is
$F^{(I)}_{\text{a-f}}$ [Eq.\ (\ref{enhancement})] with
$N^{(\textrm{a-f})}_{ccvv}=1$, and the dotted line, applicable to
carrier population and spin control is $F^{(I)}_{\text{a-a}}$ [Eq.\
(\ref{enhancement_aa})] with
$N^{(\textrm{a-a})}_{ccvv}=1$}\label{Fig_F}
\end{figure}
The Coulomb enhancement factor $F^{(I)}_{\text{a-f}}\left( \omega
\right) $ is plotted in Fig.\ \ref{Fig_F} with the approximation
that $N^{(\textrm{a-f})}_{ccvv}=1$ (see Appendix
\ref{Appx_IntStateEnhancement}).

\subsection{Carrier population control}

The `1+2' carrier population control is dominated by interference of
`allowed' one-photon transitions and `allowed-allowed' two-photon
transitions \cite{BhatSipe_unpub, Fraser03, StevensReview04}.
Substituting these into (\ref{xiSchematic}), and using the Gamma
function identity (\ref{Gamma identity}), we find
\begin{equation}
\xi _{\left( I\right) }^{ijk}=\sum_{c,v} \Xi \left( a_{cv}
\kappa_{cv} \right)\sum_{c^{\prime }v^{\prime
}}N^{(\textrm{a-a})}_{cc^{\prime }vv^{\prime }}\left( \kappa_{cv}
\right) \xi _{cc^{\prime }vv^{\prime }}^{ijk} \label{population
control result},
\end{equation}
where
\begin{equation}
\xi _{cc^{\prime }vv^{\prime }}^{ijk} = \frac{2\pi
e}{L^{3}}\sum_{\bm{\kappa }} \left( D_{cc^{\prime}vv^{\prime
}\bm{\kappa }}^{(2\text{-free})*}\right) ^{jk}\left(
D_{cv\bm{\kappa }}^{(1\text{-free})}\right) ^{l}\delta \left(
2\omega -E_{cv}\left( \kappa \right) /\hbar \right)
\end{equation}
and $\mathsf{D}_{cc^{\prime}vv^{\prime }\bm{\kappa
}}^{(2\text{-free})}$ and $\mathbf{D}_{cv\bm{\kappa
}}^{(1\text{-free})}$ are given by (\ref{D2free}) and
(\ref{D1free}). Note that only the `allowed' part of (\ref{D1free})
and the `allowed-allowed' part of (\ref{D2free}) should be retained
for a consistent solution. In the independent particle
approximation,
\begin{equation}
\xi_{\left(
I\text{-free}\right)}^{ijk}=\sum_{c^{\prime}v^{\prime}}\xi
_{cc^{\prime }vv^{\prime }}^{ijk}. \label{xi free}
\end{equation}
Thus, population control has a Coulomb enhancement due to
excitonic effects, but no phase shift.

Note that (\ref{population control result}) gives the population
control tensor at final energies above the band edge. There can
also be population control of bound excitons when both one- and
two-photon transitions are to the same excitonic state. This can
occur, for example, at $s$ excitons due to allowed-allowed
two-photon transitions \cite{Doni74} interfering with allowed
one-photon transitions.

If $N^{(\textrm{a-a})}_{cc^{\prime }vv^{\prime }}\left( \kappa_{cv}
\right)$ is approximately the same for all the terms that
significantly contribute to $\xi _{\left( I\right) }$, then, at
photon energies for which transitions from the heavy and light hole
bands dominate $\xi_{(I)}$, the Coulomb enhancement becomes
approximately independent of the sum over bands, and we can make the
simplification,
\begin{equation}
\xi _{\left( I\right) }^{ijk} \approx F^{(I)}_{\text{a-a}}
\xi_{\left( I\text{-free}\right)}^{ijk}\label{population control
simple},
\end{equation}
where
\begin{equation}
F^{(I)}_{\text{a-a}}\left( \omega \right) \equiv \Xi \left( a_{cv}
\kappa_{cv} \right) N^{(\textrm{a-a})}_{ccvv}\left( \kappa
_{cv}\right) . \label{enhancement_aa}
\end{equation}
The Coulomb enhancement factor $F^{(I)}_{\text{a-a}}\left( \omega
\right) $ is plotted in Fig.\ \ref{Fig_F} with the approximation
that $N^{(\textrm{a-a})}_{ccvv}=1$ (see Appendix
\ref{Appx_IntStateEnhancement}).

\subsection{Spin current injection and spin control}

The `1+2' spin current is dominated by interference of `allowed'
one-photon transitions and `allowed-forbidden' two-photon
transitions, whereas `1+2' spin control is dominated by interference
of `allowed' one-photon transitions and `allowed-allowed' two-photon
transitions \cite{BhatSipe_unpub}. Under the approximations that led
to (\ref{eta_C_simple}) and (\ref{population control simple}), the
spin current injection pseudotensor is
\begin{equation}
\mu_{(I)} ^{ijklm}=F^{(I)}_{\text{a-f}} \exp \left( i\delta
\right) \mu_{(I\text{-free})}^{ijklm},\label{mu}
\end{equation}
where $F^{(I)}_{\text{a-f}}$ is given by (\ref{enhancement}),
$\delta$ is given by (\ref{phase shift}), and
$\mu_{(I\text{-free})}$ is the spin current injection pseudotensor
in the independent particle approximation. Under similar
approximations, the spin control pseudotensor is
\begin{equation}
\zeta_{(I)}^{ijkl}=F^{(I)}_{\text{a-a}}
\zeta_{(I\text{-free})}^{ijkl},\label{zeta}
\end{equation}
where $F^{(I)}_{\text{a-a}}$ is given by (\ref{enhancement_aa}), and
$\zeta_{(I\text{-free})}$ is the spin control pseudotensor in the
independent particle approximation. Spin control, like carrier
population control, has a Coulomb enhancement but no phase shift.
There can also be spin control of bound excitons, but it has not
been included in (\ref{zeta}).

\section{Discussion\label{sec:discussion}}

We now examine the relationship between the Coulomb enhancements of
the `1+2' processes and of one- and two-photon absorption; the
latter are denoted by $F^{\left( 1\right) }$ and $F^{\left( 2\right)
}$ so that for $i \in \{ 1,2 \}$, $\dot{n}^{\left( i\right)
}=\dot{n}_{\text{free}}^{\left( i\right) }F^{\left( i\right) } $.
The relationship is particularly simple at photon energies for which
transitions from the heavy- and light-hole bands are dominant and
intermediate state Coulomb enhancement is the same for each
significant term in the sum over intermediate states. The Coulomb
enhancements for the `1+2' processes are then given by
(\ref{enhancement}) and (\ref{enhancement_aa}). For one-photon
absorption, $F^{\left( 1\right) } = \Xi \left( a_{cv}
\kappa_{cv}\right)$ \cite{Elliott57}. In noncentrosymmetric
semiconductors, two-photon absorption is dominated by
allowed-allowed transitions just above the band gap, and by
allowed-forbidden transitions at higher final energies; the
cross-over point in GaAs is a few meV above the band gap
\cite{vanderZiel77}. At photon energies for which allowed-allowed
transitions dominate two-photon absorption, from (\ref{two photon
a-a final}),
\begin{equation}
F^{\left( 2\right) }=\Xi \left( a_{cv} \kappa_{cv}\right) \left(
N^{(\textrm{a-a})}_{ccvv}\left( \kappa _{cv}\right) \right)^{2},
\end{equation}
and thus
\begin{equation}
F^{(I)}_{\text{a-a}}= \sqrt{ F^{\left( 1\right) } F^{\left( 2\right)
} } \text{ and } F^{(I)}_{\text{a-f}}= C\sqrt{ F^{\left( 1\right) }
F^{\left( 2\right) } },\label{F_geometric_mean_aa}
\end{equation}
where $C \equiv \left( N^{(\textrm{a-f})}_{ccvv}\left( \kappa
_{cv}\right) / N^{(\textrm{a-a})}_{ccvv}\left( \kappa
_{cv}\right)\right) \sqrt{1+\left( a_{cv}\kappa_{cv} \right)
^{-2}}$, while at photon energies for which allowed-forbidden
transitions dominate two-photon absorption, from (\ref{two photon
a-f}),
\begin{equation}
F^{\left( 2\right) }=\Xi \left( a_{cv} \kappa_{cv}\right)\left(
1+\left( a_{cv} \kappa_{cv} \right) ^{-2}\right) \left(
N^{(\textrm{a-f})}_{ccvv}\left( \kappa _{cv}\right)\right)^{2},
\label{F2 af}
\end{equation}
and thus
\begin{equation}
F^{(I)}_{\text{a-a}}= \left( 1/C \right) \sqrt{ F^{\left( 1\right) }
F^{\left( 2\right) } } \text{ and } F^{(I)}_{\text{a-f}}= \sqrt{
F^{\left( 1\right) } F^{\left( 2\right) }
}.\label{F_geometric_mean_af}
\end{equation}
Note that, based on Appendix \ref{Appx_IntStateEnhancement}, $C
\approx \sqrt{1+B_{cv}/\left(  2\hbar \omega - E_{g} \right) }$,
which is the ratio of the two curves in Fig.\ \ref{Fig_F}. In
centrosymmetric semiconductors, there are no allowed-allowed
transitions, and only (\ref{F_geometric_mean_af}) applies.

The `1+2' processes are often described by ratios. For example, a
useful quantity to describe the current is the swarm velocity
\cite{HacheIEEE98,SipeShkrebtiiPulci98}, defined as the average
velocity per injected electron-hole pair
\[
\mathbf{v}_{\text{swarm}}\equiv \frac{\left( d\mathbf{J}/dt\right) }{e\left(
dn/dt\right) }.
\]
The swarm velocity is a maximum when the relative intensities of
the two colors are chosen such that $\dot{n}_{2\omega
}=\dot{n}_{\omega }$; returning to (\ref{currentMicro}), if one
associates the one- and two-photon amplitudes with the arms of an
effective interferometer, this condition corresponds to balancing
that interferometer. For fields co-linearly polarized along
$\mathbf{\hat{x}}$, the maximum swarm speed is
\begin{equation}
v_{\text{swarm}}=\frac{1}{e}\frac{\left| \eta_{(I)} ^{xxxx}\right|
}{\sqrt{\xi _{\left( 1\right) }^{xx}\xi _{\left( 2\right)
}^{xxxx}}} . \label{swarm expression}
\end{equation}
A useful quantity to describe pure spin currents is the maximum spin
separation distance \cite{HubnerPRL03}; it is proportional to
$\mu_{(I)}/\sqrt{\xi _{(1)}\xi_{(2)}}$. As a consequence of
(\ref{F_geometric_mean_af}), the maximum swarm speed, and the
maximum spin separation distance, are \emph{unaffected} by excitonic
effects when allowed-forbidden transitions dominate two-photon
absorption \cite{footnote:SpinSeparationDistance}. However, close to
the band edge, where allowed-allowed transitions dominate two-photon
absorption, excitonic effects increase these ratios by a factor $C$
over their value in the independent particle approximation. In
contrast, as a consequence of (\ref{F_geometric_mean_af}), excitonic
effects do not affect the maximum control ratio for population and
spin control ($\xi_{(I)}/\sqrt{\xi _{(1)}\xi_{(2)}}$ and
$\zeta_{(I)}/\sqrt{\xi _{(1)}\xi_{(2)}}$ respectively
\cite{Fraser99PRL, Stevens_pssb, StevensReview04}) close to the band
edge and decrease them by a factor $C$ at higher photon energies for
which allowed-forbidden transitions dominate two-photon absorption.

In the terminology of Seideman \cite{Seideman98}, the excitonic
phase shift of the `1+2' current and spin current is a direct phase
shift. It is due to the complex nature of the final state as it
appears in the transition amplitudes. Thus it can be understood in
terms of the partial wave phase shifts of the final state caused by
the Coulomb potential between electron and hole. The Coulomb
interaction is rather unique due to its long range nature, so we
first suppose the potential between the electron and hole falls off
more rapidly than $1/C$. In that simpler problem, the final state
wave function is written as
\[
\psi _{\bm{\kappa }}(\mathbf{r})=\sum_{l=0}^{\infty
}i^{l}e^{-i\delta_{l}\left(\kappa \right)}\left( 2l+1\right)
\frac{u_{\kappa ,l}\left( r\right) }{r} P_{l}\left(
\frac{\mathbf{r}\cdot \bm{\kappa }}{r\kappa }\right) ,
\]
where the $u_{\kappa ,l}\left( r\right) $ are real
\cite{TaylorScatteringTheory}. If the potential between the
particles is ignored, then the partial wave phase shifts, $\delta
_{l}\left(\kappa \right)$ are zero. The allowed one-photon pathway
reaches an $s$ wave, while the allowed-forbidden two-photon pathway
reaches a $p$ wave. Substituting this form for the wave function
into the one- and two-photon transition amplitudes, yields $\Omega
_{\bm{\kappa }}^{(1)}=\Omega _{\bm{\kappa
}}^{(1\text{-free})}e^{i\delta _{0}\left(\kappa
\right)}f_{0}\left(\kappa \right)$ for the one-photon rate, where
$f_{0}\left(\kappa \right)$ is real and depends on $u_{\kappa
,0}\left( r\right) $, and $\Omega _{\bm{\kappa
}}^{(2\text{:a-f})}=\Omega _{\bm{\kappa
}}^{(2\text{-free})}e^{i\delta _{1}\left(\kappa
\right)}f_{1}\left(\kappa \right)$ for the two-photon rate, where
$f_{1}\left(\kappa \right)$ is real and depends on $u_{\kappa
,0}\left( r\right) $ and $u_{\kappa ,1}\left( r\right) $. Here
$\Omega _{\bm{\kappa }}^{(i\text{-free})} $ is the \textit{i}-photon
transition amplitude when the potential between the particles is
ignored. It is then straightforward to see from (\ref{etaSchematic})
that the relative shift of the partial waves is responsible for the
phase shift of the current and spin current. That is,
\begin{equation}
\delta =\delta _{0}-\delta _{1}.  \label{related to partial waves}
\end{equation}
The use of ionization states as opposed to scattering states was
important to get the correct sign of the intrinsic phase. With
scattering states, one would find $\delta=\delta_{1} - \delta_{0}$.

Due to the long-range nature of the Coulomb potential, the partial
wave phase shifts have a logarithmic $r$ dependent part, but it is
the same for all partial waves and thus does not appear in the
relative phase. The part of the Coulomb partial wave phase shift
$\delta_{l}\left(\kappa \right)$ that does not depend on $r$ is
$\arg \left(\Gamma \left( l+1+i/(a_{cv}\kappa )\right) \right)$
\cite{BetheSalpeter, TaylorScatteringTheory}; when inserted into
(\ref{related to partial waves}), this reproduces (\ref{phase
shift}). In contrast, the allowed-allowed two-photon pathway reaches
an $s$ wave and thus there is no phase shift for population control
or spin control.

The expression (\ref{related to partial waves}) for the intrinsic
phase in terms of the scattering phases is particularly simple,
since each pathway connects to only a single parity. This contrasts
with `1+2' ionization from an atomic $s$ state, for which the
one-photon transition is to a $p$ wave and the two-photon transition
is to both $s$ and $d$ waves; the intrinsic phase is thus a
weighting of the $p$-$s$ and $p$-$d$ partial wave shifts
\cite{Baranova90}. Materials for which the first term in
(\ref{Bloch_v_expansion}) is forbidden (Cu$_{2}$O is an example)
have these same selection rules \cite{Elliott57, Mahan68,
Rustagi73}; hence, they will have an intrinsic phase with a similar
weighting.

The absence of a phase shift in population control can be
connected to a symmetry of the second order nonlinear optical
susceptibility. From considerations of energy transfer and
macroscopic electrodynamics, $\xi_{(I)}$ is related to the
nonlinear susceptibility $\chi^{(2)}$ by
\begin{equation}\label{eqn:ConnectingToSusceptibilities}
    \xi_{(I)}^{ijk}=\left( i\epsilon_{0}/\hbar\right)\left[
    \chi^{(2)kij}\left(2\omega; -\omega,-\omega\right)
    -\chi^{(2)jki}\left(-\omega; 2\omega,-\omega\right)
    \right].
\end{equation}
In the independent particle approximation \cite{SipeShkrebtii00},
\begin{equation}\label{GOPR}
   \chi ^{\left( 2\right) ijk}\left( 2\omega ;-\omega ,-\omega \right)
    =\left[ \chi ^{\left( 2\right)jik}\left( -\omega ;2\omega ,-\omega \right)\right]
    ^{*},
\end{equation}
which is a generalization of overall permutation symmetry to
resonant absorption. As a result of (\ref{GOPR}), Fraser \textit{et
al}.\ showed that $\xi_{\left( I\right) }$ is proportional to
$\mathrm{Im} \chi^{(2)}$, and is thus purely real
\cite{Fraser99PRL}. Our result that $\xi_{(I)}$ remains real when
excitonic effects are included suggests that (\ref{GOPR}) holds more
generally. In fact, it can be shown that (\ref{GOPR}) holds for any
Hamiltonian symmetric under time-reversal so long as $\hbar \omega $
is not resonant.

\section{Summary and Outlook}

We have extended the theory of interband `1+2' processes in bulk
semiconductors to include the electron-hole interaction. Following
previous theories
\cite{Atanasov96,Fraser99PRL,BhatSipe00,Najmaie03,Stevens_pssb}, we
have used a framework based on (i) a separation of the initial
carrier photoinjection and the subsequent carrier scattering, and
(ii) a perturbative expansion in the optical field amplitudes, with
injection rates obtained in a Fermi's golden rule limit for the
bichromatic field. The injection rates for carrier population
control, spin control, current injection, and spin current
injection, have been described phenomenologically by tensors
$\xi_{(I)}$, $\zeta_{(I)}$, $\eta_{(I)}$, and $\mu_{(I)}$,
respectively
\cite{Atanasov96,Fraser99PRL,Najmaie03,Stevens_pssb,StevensReview04}.
Like previous theories, we have used the long-wavelength limit, and
neglect nonlocal corrections to the interaction Hamiltonian. But
whereas previous theories of `1+2' photoinjection used the
independent particle approximation, we have included excitonic
effects. We have shown that excitonic effects cause (i) an
enhancement of each `1+2' process, and (ii) a phase shift for
current injection and spin current injection. Our main results, the
modifications of the aforementioned tensors relative to the
independent particle approximation are given in
(\ref{eta_C_simple}), (\ref{population control simple}), (\ref{mu}),
and (\ref{zeta}). These particularly simple results are valid at
photon energies for which transitions from the heavy- and light-hole
bands are dominant; more general results are given for $\eta_{(I)}$
and $\xi_{(I)}$ in (\ref{eta_result}) and (\ref{population control
result}).

Our results are based on the effective mass model of Wannier
excitons; degenerate bands are included, but we use a spherical
approximation to the exciton Hamiltonian, and we neglect
envelope-hole coupling. This is a good approximation for many
typical semiconductors, including GaAs, since the electron-hole
envelope function extends over many unit cells due to the screening
of the Coulomb interaction by the static dielectric constant
\cite{Baldereschi70, Baldereschi71, Baldereschi73, Sondergeld77I,
Sondergeld77II}. As a consequence of making the spherical
approximation, the phase shifts and Coulomb enhancements we find in
this paper are independent of crystal orientation.

Also, our results are limited to low excess photon energy since (i)
the Wannier exciton Hamiltonian assumes parabolic Bloch bands, and
(ii) we have truncated the expansion in $\mathbf{k}$ of the Bloch
state velocity matrix elements, which is the basis of the transition
amplitude expansion. By comparing the black dashed line and grey
dash-dotted line in Fig.\ \ref{Fig_eta}, one sees that higher order
terms in $\mathbf{k}$ (for both bands and velocities) are important
in GaAs for excess photon energies greater than about 200 meV. This
can then be considered the limit of validity of our calculation.
However, combining the Coulomb enhancement calculated assuming
parabolic bands with the nonperturbative independent particle
approximation result (as was done for the solid black line in Fig.\
\ref{Fig_eta}) likely gives a good approximation for a few hundred
more meV; this was the case for one- and two-photon absorption
\cite{Sturge62,Weiler81}.

It is interesting to ask if there are other sources of intrinsic
phases to the current (or spin current) besides the one that we have
identified here, as these may produce spectral features in the
intrinsic phase. One possibility is the coupling between bound
$so$-$c$ excitons and the unbound $hh$-$c$ or $lh$-$c$ excitons,
since it is known that the intrinsic phase can show spectral
features near a resonance \cite{Seideman99}. Another possibility is
the envelope-hole coupling between the continua of unbound $hh$-$c$
and $lh$-$c$ excitons that was neglected in our treatment. The
effect of coupled continua in the general theory of `$n+m$' phase
shifts has not been considered to date.

Finally, we note that the intrinsic phase and Coulomb enhancement
may be greater in reduced dimensional systems, which have greater
exciton binding energies. The carrier-carrier Coulomb interaction
was included in the theory for `$1+2$' control of electrons in
biased asymmetric quantum wells, although the intrinsic phase was
not studied \cite{Potz98}.

\appendix

\section{Intermediate state Coulomb enhancement \label{Appx_IntStateEnhancement}}

Consider the functions $N^{(\textrm{a-f})}$ and
$N^{(\textrm{a-a})}$, which appear in (\ref{N_af}) and (\ref{N_aa});
we refer to them collectively as $N$. First, note that due to the
energy conserving delta function in (\ref{etaSchematic}), $\kappa$
will be equal to $\kappa_{cv}$ [see Eq.\ (\ref{energy
conservation})], and thus $N$ is a function only of $\omega $,
$E_{cv}^{g}$, $E_{c^{\prime }v^{\prime }}^{g}$, $\mu_{cv}$, and
$\mu_{c^{\prime}v^{\prime}}$. Second, note that $N$ is defined so
that if the electron-hole attraction is turned off, for example by
letting $\epsilon \rightarrow \infty $, then $N\rightarrow 1$
\cite{footnote:Mahan_J}. This allows $N$ to be identified as part of
the Coulomb enhancement. In particular, $N$ is the enhancement due
to the Coulomb interaction in the intermediate states; if the
Coulomb interaction is neglected for the intermediate states, $N=1$
\cite{Doni74}. [Note that Lee and Fan \cite{LeeFan74} did not allow
for $v^{\prime }\neq v$ in $N$ (related to $J_{j}$ in their
notation).]

\begin{figure}
\includegraphics[width=3.2in]{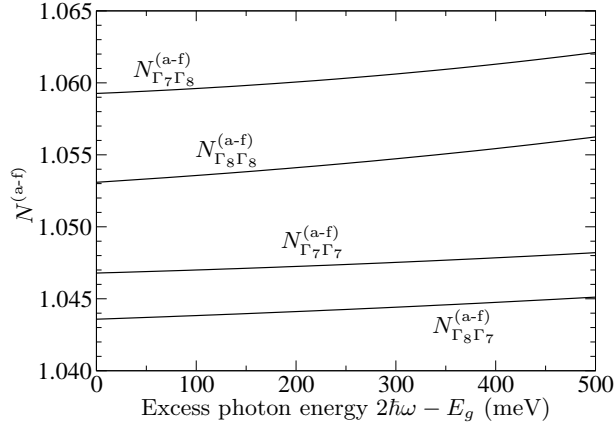}
\caption{The factor $N^{(\textrm{a-f})}$ [intermediate state Coulomb
enhancement, see (\ref{two photon a-f})] for GaAs. The first
(second) subscript is $v$ ($v^{\prime}$); subscripts for
$c=c^{\prime}=\Gamma_{6}$ are not shown; $\Gamma _{8}$ denotes the
heavy and light hole bands, and $\Gamma _{7}$ denotes the split off
band.} \label{Fig_N}
\end{figure}

Since the integrand is smooth for the parameter range of interest,
numerical integration of $N$ is straightforward; however, it need
not be undertaken. Further simplification is possible since the
parameter $\gamma $ can be considered to be much less than one.
Since most materials have an exciton binding energy that is much
smaller than the band gap, $\hbar \omega $ is detuned from the band
edge by many exciton binding energies at photon energies consistent
with the approximations made here. In GaAs, for example, when
$2\hbar \omega $ is within 500\,meV of the gap, $\gamma $ is at most
$0.09$. An expansion of $N^{(\textrm{a-f})}$ for small $\gamma $,
\[
N^{(\textrm{a-f})}=1+\frac{2}{3}\gamma _{c^{\prime }v^{\prime
}}+\left( \frac{4}{3}\ln 2-\frac{1}{3}\right) \gamma _{c^{\prime
}v^{\prime }}^{2}+\left( S_{0}- \frac{2}{15}a_{cv}^{2}\kappa
^{2}\right) \gamma _{c^{\prime }v^{\prime }}^{3}+O\left( \gamma
_{c^{\prime }v^{\prime }}^{4}\right),
\]
where $S_{0}\approx 0.5633 $, shows that $N^{(\textrm{a-f})}$ is
approximately $1$ and nearly constant as a function of $\omega $.
The same is true of $N^{(\textrm{a-a})}$, which has the expansion
\[
N^{(\textrm{a-a})}=1-\frac{2}{a_{cv}}\left( a_{c^{\prime }v^{\prime
}}-a_{cv}\right) P +O\left( \gamma _{c^{\prime }v^{\prime
}}^{4}\right),
\]
where, with $S_{1}\approx 1.645$, \[ P \equiv \gamma _{c^{\prime
}v^{\prime }}+\left( 2\ln 2 - \frac{a_{c^{\prime }v^{\prime
}}}{a_{cv}}\right) \gamma _{c^{\prime }v^{\prime }}^{2} + \left(
\frac{2a_{c^{\prime }v^{\prime}}^{2}}{3a_{cv}^{2}} -
\frac{2a_{c^{\prime }v^{\prime }}}{a_{cv}} - \frac{1}{3}a_{c^{\prime
}v^{\prime }}^{2}\kappa ^{2} + S_{1}\right) \gamma _{c^{\prime
}v^{\prime }}^{3}. \] In fact, when
$\mu_{cv}=\mu_{c^{\prime}v^{\prime}}$, $N^{(\textrm{a-a})}=1$ even
to fourth order in $\gamma_{c^{\prime }v^{\prime }}$. Fig.\
\ref{Fig_N} shows a numerical integration of $N^{(\textrm{a-f})}$
using the parameters of GaAs.

\section{Evaluation of the current injection tensor \label{Appx_etaFree}}

The tensor $\eta _{cc^{\prime}vv^{\prime}}^{ijkl}$, defined in
(\ref{eta_ccvv}), can be used to calculate the current injection
tensor with or without excitonic effects using (\ref{eta_result}) or
(\ref{eta_result_free}). It can be evaluated analytically in the
approximation of parabolic bands. Part of the result for the 8 band
Kane model has been given before but without the split-off band as
an initial or intermediate state \cite{BhatSipe00}. We here give
more detail, but only for the two-band terms. We denote the bands by
a double index $n$ and $s$, where $n$ is one of $\left\{
c,hh,lh,so\right\} $, and $s$ runs over the two spin states for each
band. Since the Coulomb corrections to $\eta_{(I)}$ in
(\ref{eta_result}) do not depend on the spin index, we can include
the sum over spin indices from (\ref{eta_result}) in $\eta
_{c,v,v^{\prime }}^{ijkl}$. And, since we are only calculating
two-band terms, we set $c^{\prime }=c$ and $v^{\prime }=v$. Thus
\begin{eqnarray*}
\eta _{ccvv}^{ijkl} &=&i\frac{\pi e^{4}}{\hbar ^{2}\omega ^{3}}
\sum_{s,s^{\prime }}\frac{1}{L^{3}}\sum_{\mathbf{k}}\Delta
_{cv}^{i}\frac{\left\{ \left(
\mathbf{v}_{cc}-\mathbf{v}_{vv}\right) ,\mathbf{v} _{cs,vs^{\prime
}}^{*}\right\} ^{jk}}{E_{cv}-\hbar \omega }v_{cs,vs^{\prime
}}^{l}\delta \left( 2\omega -E_{cv}\left( k\right) /\hbar \right) \\
&=&i\frac{\pi e^{4}}{\hbar ^{3}\omega ^{4}}\frac{1}{8\pi ^{3}}\int
k^{2}dk \frac{\mu_{cv}}{\hbar k_{cv}}\delta \left( k-k_{cv}\right)
d\Omega \sum_{s,s^{\prime }}\frac{\hbar
k^{i}}{\mu_{cv}}\frac{1}{2}\frac{\hbar k^{j}}{
\mu_{cv}}v_{vs^{\prime },cs}^{k}v_{cs,vs^{\prime }}^{l}+\left(
j\leftrightarrow k\right) \\
&=&i\frac{e^{4}}{\hbar ^{2}\omega ^{4}}\frac{1}{8\pi
^{2}}\frac{k_{cv}^{3}}{ \mu_{cv}}\frac{1}{2}\int d\Omega
\hat{k}^{i}\hat{k}^{j}\sum_{s,s^{\prime }}v_{vs^{\prime
},cs}^{k}v_{cs,vs^{\prime }}^{l}+\left( j\leftrightarrow k\right)
\end{eqnarray*}
where
\[
k_{cv}=\sqrt{\frac{2\mu_{cv}}{\hbar ^{2}}\left( 2\hbar \omega
-E^{g}_{cv}\right) }.
\]
For $v=lh$ or $hh$, $E_{cv}^{g}=E_{g}$, while for $v=so$,
$E_{cv}^{g}=E_{g}+\Delta $, where $E_{g}$ is the fundamental band
gap, and $\Delta $ is the spin-orbit splitting.  The interband
velocity matrix elements are approximated by their value at $k=0$,
but still depend on the direction of $\mathbf{k}$. In terms of the
orthogonal triple of unit vectors $\mathbf{\hat{k}}$,
$\mathbf{\hat{l}}$ and $\mathbf{\hat{m}}$,
\begin{eqnarray*}
\mathbf{\hat{k}} &=&\sin \theta \cos \phi \mathbf{\hat{x}}+\sin
\theta \sin \phi \mathbf{\hat{y}}+\cos \theta \mathbf{\hat{z}}
\\
\mathbf{\hat{l}} &=&\cos \theta \cos \phi \mathbf{\hat{x}}+\cos
\theta \sin
\phi \mathbf{\hat{y}}-\sin \theta \mathbf{\hat{z}} \\
\mathbf{\hat{m}} &=&-\sin \phi \mathbf{\hat{x}}+\cos \phi
\mathbf{\hat{y}},
\end{eqnarray*}
these matrix elements are:
\begin{eqnarray*}
\mathbf{v}_{c,s;hh,s^{\prime }} &=&\frac{1}{2}\sqrt{\frac{E_{P}}{m}}\left[
\mathbf{\hat{l}}\mathbb{I}+i\mathbf{\hat{m}}\sigma _{z}\right] _{s,s^{\prime
}} \\
\mathbf{v}_{c,s;lh,s^{\prime }}
&=&-\frac{1}{2}\sqrt{\frac{E_{P}}{3m}} \left[
2\mathbf{\hat{k}}\mathbb{I}+i\mathbf{\hat{l}}\sigma
_{y}-i\mathbf{\hat{m} }\sigma _{x}\right] _{s,s^{\prime }} \\
\mathbf{v}_{c,s;so,s^{\prime }} &=& \sqrt{\frac{E_{P}}{6m}}\left[
\mathbf{\hat{k}}\mathbb{I}-i\mathbf{\hat{l}}
\sigma_{y}+i\mathbf{\hat{m}}\sigma_{x}\right]_{s,s^{\prime }},
\end{eqnarray*}
where $E_{P}$ is the Kane energy \cite{Kane57}. Here, $\mathbb{I}$
is the $2\times 2$ identity matrix and $\sigma _{i}$ are the Pauli
spin matrices. Of course, for parabolic bands, the intraband
matrix elements are $\left\langle n,s, \mathbf{k}\right|
\mathbf{v}\left| n,s^{\prime },\mathbf{k} \right\rangle =\delta
_{s,s^{\prime }}\mathbf{\hat{k}}\hbar k/m_{n}$, where $m_{n}$ is
the effective mass of band $n$. (In the proper Kane model, the
effective masses are given in terms of the parameters $E_{g}$,
$\Delta $, and $E_{P}$, but we treat them as additional
parameters, which is equivalent to including remote band effects
on the effective masses.) The sums over spin then yield
\begin{eqnarray*}
\sum_{s,s^{\prime }}v_{hh,s^{\prime };c,s}^{k}v_{c,s;hh,s^{\prime }}^{l} &=&
\frac{E_{P}}{2m}\left( \delta _{k,l}-\hat{k}^{k}\hat{k}^{l}\right) \\
\sum_{s,s^{\prime }}v_{lh,s^{\prime };c,s}^{k}v_{c,s;lh,s^{\prime
}}^{l} &=& \frac{E_{P}}{2m}\left(
\hat{k}^{k}\hat{k}^{l}+\frac{1}{3}\delta _{k,l}\right) \\
\sum_{s,s^{\prime }}v_{so,s^{\prime };c,s}^{k}v_{c,s;so,s^{\prime
}}^{l} &=&\frac{E_{P}}{3m}\delta _{k,l}.
\end{eqnarray*}
The remaining angular integrals can be done using
\begin{eqnarray*}
\int d\Omega \hat{k}^{i}\hat{k}^{j} &=&\frac{4\pi }{3}\delta _{i,j} \\
\int d\Omega \hat{k}^{i}\hat{k}^{j}\hat{k}^{k}\hat{k}^{l}
&=&\frac{4\pi }{15} \left( \delta _{i,j}\delta _{k,l}+\delta
_{i,k}\delta _{j,l}+\delta _{i,l}\delta _{j,k}\right).
\end{eqnarray*}

The result for $\eta_{(I\text{-free})}$ is
\begin{equation}
\eta_{(I\text{-free})}^{ijkl}=i\frac{\sqrt{2}}{9\pi
}\frac{e^{4}E_{P}}{\omega ^{4}\hbar ^{5}\sqrt{m}} \left[ \left(
2\hbar \omega -E_{g}\right) ^{\frac{3}{2}
}\left(T_{hh}^{ijkl}+T_{lh}^{ijkl} \right) + \left( 2\hbar \omega
-E_{g} - \Delta \right) ^{\frac{3}{2}} T_{so}^{ijkl} \right],
\label{etaFree parabolic}
\end{equation}
where the tensor properties are
\begin{eqnarray*}
T_{hh}^{ijkl} &=&\sqrt{\frac{\mu_{c,hh}}{m}}\left(
\frac{9}{20}\delta _{i,j}\delta _{k,l}+\frac{9}{20}\delta
_{i,k}\delta _{j,l}-\frac{3}{10}\delta
_{i,l}\delta _{j,k}\right) \\
T_{lh}^{ijkl} &=&\sqrt{\frac{\mu_{c,lh}}{m}}\left(
\frac{11}{20}\delta _{i,j}\delta _{k,l}+\frac{11}{20}\delta
_{i,k}\delta _{j,l}+\frac{3}{10}\delta _{i,l}\delta _{j,k}\right)
\\
T_{so}^{ijkl} &=& \frac{1}{2}\sqrt{\frac{\mu_{c,so}}{m}}\left(
\delta _{i,j}\delta _{k,l}+\delta _{i,k}\delta _{j,l}\right),
\end{eqnarray*}
and the term involving $T_{so}^{ijkl}$ should not be included if
$2\hbar \omega < E_g + \Delta$. The free particle current
injection tensor has also been investigated for parabolic bands,
but with a simple three-band model \cite{Sheik-Bahae99}. That
model does not have the matrix elements $\mathbf{v}_{c\uparrow
,lh\downarrow }$ and $\mathbf{v}_{c\downarrow ,lh\uparrow }$, and
thus differs from our result for $T_{lh}^{xxxx}$.

\begin{acknowledgments}
This work was financially supported by the Natural Science and
Engineering Research Council, Photonics Research Ontario, and the
US Defense Advanced Research Projects Agency. We gratefully
acknowledge many stimulating discussions with Daniel C\^{o}t\'{e},
James Fraser, Ali Najmaie, Fred Nastos, Eugene Sherman, Art Smirl,
Marty Stevens, and Henry van Driel.
\end{acknowledgments}


\begin{thebibliography}{93}
\bibitem[{foo({\natexlab{b}})}]{footnote:SpinSeparationDistance}
\bibinfo{note}{At higher excitation densities, the Coulomb interaction can
  modify the maximum spin separation distance through the momentum relaxation
  time}.
\end{thebibliography}
\end{document}